\RequirePackage{etoolbox}
\newbool{fullversion}
\booltrue{fullversion}
\pdfoutput=1
\RequirePackage{etoolbox}

\providebool{fullversion}
\providebool{arxivversion}

\providebool{diffversion}

\boolfalse{diffversion}

\newcommand{\diffred}[1]{\ifbool{diffversion}
    {\textcolor{red}{#1}}
    {#1}}

\documentclass[acmsmall,%
  nonacm
]{acmart}%

\usepackage{preamble}
\usepackage{iris}
\usepackage{prob}
\usepackage{examples}

\ifbool{fullversion}{
  \newcommand{\appref}[1]{\cref{#1}}
  \newcommand{\Appref}[1]{\Cref{#1}}
}{
  \newcommand{\appref}[1]{the extended version of this paper~\cite{foxtrot-extended}}
  \newcommand{\Appref}[1]{the extended version of this paper~\cite{foxtrot-extended}}
}

\ifbool{arxivversion}{
\usepackage{caption}
}{
\usepackage[belowskip=-10pt,aboveskip=4pt]{caption}
\setlength{\intextsep}{15pt} %
}

\makeatletter
\ifbool{arxivversion}{}{
\let \MathparLineskip \mpr@lesslineskip %
}
\makeatother

\ifbool{arxivversion}{
  \newcommand{\vsquish}[1]{}
}{
  \newcommand{\vsquish}[1]{\vspace{-#1}}
}

\usepackage{tcolorbox}
\tcbuselibrary{skins,breakable}
  \newtcolorbox{result}{
    blanker,
    extras={interior engine=spartan},
    grow to left by=2pt,left*=0mm,
    grow to right by=2pt,right*=0mm,
    top=1mm,bottom=1mm,
    beforeafter skip balanced=0.1\baselineskip plus 2pt,
    breakable,
    colback=cyan!40
  }

\ifbool{fullversion}{
}{}

\setcopyright{cc}
\setcctype{by}
\acmDOI{10.1145/3808265}
\acmYear{2026}
\acmJournal{PACMPL}
\acmVolume{10}
\acmNumber{PLDI}
\acmArticle{187}
\acmMonth{6}
\acmSubmissionID{pldi26main-p77-p}
\received{2025-11-10}
\received[accepted]{2026-04-03}

\begin{document}



\bibliographystyle{ACM-Reference-Format}

\title{Contextual Refinement of Higher-Order Concurrent Probabilistic Programs  \ifbool{fullversion}{(Extended Version)}{}}

\author[K. H. Li]{Kwing Hei Li}
\orcid{0000-0002-4124-5720}
\affiliation{%
  \institution{Aarhus University}
  \city{Aarhus}
  \country{Denmark}
}
\email{hei.li@cs.au.dk}

\author[A. Aguirre]{Alejandro Aguirre}
\orcid{0000-0001-6746-2734}
\affiliation{%
  \institution{Aarhus University}
  \city{Aarhus}
  \country{Denmark}
}
\email{alejandro@cs.au.dk}

\author[J. Tassarotti]{Joseph Tassarotti}
\orcid{0000-0001-5692-3347}
\affiliation{%
  \institution{New York University}
  \city{New York}
  \country{USA}
}
\email{jt4767@nyu.edu}
\authornote{Also affiliated with Amazon Web Services. This paper does not reflect the views of Amazon Web Services.}

\author[L. Birkedal]{Lars Birkedal}
\orcid{0000-0003-1320-0098}
\affiliation{%
  \institution{Aarhus University}
  \city{Aarhus}
  \country{Denmark}
}
\email{birkedal@cs.au.dk}

\begin{abstract}
  We present Foxtrot, the first higher-order separation logic for proving contextual refinement of higher-order concurrent probabilistic programs with higher-order local state.
  From a high level, Foxtrot inherits various  concurrency reasoning principles from standard concurrent separation logic, e.g. invariants and ghost resources, and supports advanced probabilistic reasoning principles for reasoning about complex probability distributions induced by concurrent threads, e.g. tape presampling and induction by error amplification. 
  The integration of these strong reasoning principles is highly \emph{non-trivial} due to the combination of probability and concurrency in the language and the complexity of the Foxtrot model; the soundness of the logic relies on a version of the axiom of choice within the Iris logic, which is not used in earlier work on Iris-based logics.
  We demonstrate the expressiveness of Foxtrot on a wide range of examples, including the adversarial von Neumann coin and the $\mathsf{randombytes\_uniform}$ function of the Sodium cryptography software library.
  
  All  results have been mechanized in the Rocq proof assistant and the Iris separation logic framework.

\end{abstract}

\maketitle

\section{Introduction}
\label{sec:introduction}
Randomization and concurrency are, independently, two important and widely-used
language features in the world of computing. The former is used to implement
 probabilistic data structures, randomized algorithms, and machine
learning algorithms; the latter is utilized to increase program throughput and
to model distributed systems and multi-core architecture. Together, the
combination of these two features can be found in concurrent randomized
algorithms~\cite{morris, concurrentbst} and in cryptography
protocols~\cite{securempc}.
However, formal methods for reasoning about concurrent probabilistic programs are limited and not very
expressive, in part because concurrency introduces nondeterminism, which in semantic models does not
compose straightforwardly with probabilistic choice \cite{varacca}.

\subsection{Motivating Example and Prior Work}
\label{sec:motivating-example}
\newcommand{\fimpl}{\textlog{fImpl}}
\newcommand{\fspec}{\textlog{fSpec}}
Consider the following $\fimpl$ function, where $\Rand\tapebound$ samples an integer uniformly from $\{0,\dots,\tapebound\}$:
\smalldisplay{
  \begin{align*}
  \fimpl\eqdef{} &  \Lam \_. \Let (x,y) = (\Parallel{\Rand 7}{\Rand31 }) in  %
  \ x\cdot 32 + y
\end{align*}}%
Though $\fimpl$ appears simple, it already exhibits the combination of both probability and concurrency. 
When we call $\fimpl$, it spawns two concurrent threads, which generate a random $3$-bit and $5$-bit number, respectively. Afterwards, it returns the binary concatenation of the two numbers generated.
 Note that since the random samples of both threads are independent, this function behaves the same as one that directly samples a random $8$-bit number. Hence to reason about its correctness, one might want to show that its behavior is the same  as the following $\fspec$ function.
 \smalldisplay{
   \begin{align*}
  \fspec \eqdef{} &\Lam \_. \Rand  255
\end{align*}}%
To formally argue that the two programs behave the same, we might want to prove that $\fimpl$ is \emph{contextually equivalent} to  $\fspec$. 
Intuitively, an expression $\expr_{1}$ is contextually equivalent to another expression $\expr_{2}$ at
a type $\type$ if no well-typed contexts $\ctx$ of ground type can distinguish them, denoted by $\ctxeq{\emptyset}{e_1}{e_2}{\type}$. In other words,  the behavior of a client program remains unchanged if we replace any occurrence of the sub-program $\expr_1$ with $\expr_2$. 
Contextual equivalence is defined as the symmetric interior of contextual refinement,
denoted by $\ctxrefines{\emptyset}{e_1}{e_2}{\type}$, which is roughly defined to
mean that, for any context $\ctx$ the observable behaviour of $\fillctx\ctx[\expr_{1}]$
is \emph{included} in the observable behaviour of $\fillctx \ctx[\expr_{2}]$.  Proving contextual equivalence or refinement is generally a  difficult problem as the statement quantifies over \emph{all} possible contexts, including those that might use concurrency and probability as well.
\diffred{Although there are other techniques for proving contextual equivalence or refinement of concurrent programs~\cite{reloc} or probabilistic programs~\cite{wandLR,probabilityLR}, respectively, to the best of our knowledge, there is \emph{no} prior work on verifying contextual equivalence of higher-order programs utilizing \emph{both} probability and concurrency.
(See \cref{fig:relwork} for an overview of a selection of prior work for proving contextual refinement of less expressive languages and \cref{sec:related-work} for a more detailed summary. \Cref{fig:relwork} also compares these techniques on whether they are implemented as a modular program logic, support reasoning about programs with local state, and are mechanized in a proof assistant.)}

\begin{figure}[h]%
\newcommand\no{\Circle}
\newcommand\kindof[1]{\LEFTcircle\ \ (#1)}
\newcommand\yes{\CIRCLE}
  \centering\small%
  \begin{tabular}{m{4cm} c c c c c}
    &  \textbf{Probability}&\textbf{Concurrency} & \textbf{Logic}  &   \textbf{Local State}  & \textbf{Mechanized} \\
    \textbf{Prob. $\lambda$-calculi~\cite{cbneed-ce, cbname-ce, cbv-ce}} &\yes&\no & \no  &\no & \no \\
    \textbf{\citet{wandLR}} &  \yes  &\no & \no & \no & \no \\
    \textbf{\citet{probabilityLR}} &  \yes &\no & \no& \yes & \no \\
    \textbf{ReLoC~\cite{reloc}} &  \no & \yes &\yes &\yes& \yes \\
    \textbf{\citet{prob-nondeterminismLR}} &  \yes &\kindof{ND} & \no & \no& \no \\
    \textbf{Clutch~\cite{clutch}} &  \yes & \no&\yes & \yes  & \yes\\
    \textbf{Approxis~\cite{approxis}} &  \yes & \no& \yes& \yes   & \yes\\
    \textbf{\theaplog (this paper)} & \yes & \yes  & \yes & \yes & \yes
  \end{tabular}
  \caption{Systems for proving contextual refinement (Logic = implemented as a logic,  ND = only nondeterminism).}
  \label{fig:relwork}
\end{figure}

\subsection{Contributions and Key Ideas}
\label{sec:key-ideas}

In this paper, we present \emph{Foxtrot}, a higher-order probabilistic
concurrent separation logic, which we use to build
the \emph{first} logical relations model for proving contextual refinement and
equivalence of concurrent probabilistic programs. At a high level, the Foxtrot
logic supports reasoning about contextual refinement of higher-order  probabilistic
concurrent imperative programs written in \thelang, an ML-style
programming language with higher-order dynamically-allocated local state,
discrete probabilistic sampling, and dynamically allocated concurrently running
threads.

Since \thelang is probabilistic, the observable behaviour of a program refers to the \emph{probability}
of observing some behaviour, such as termination.  Hence, by contextual
refinement $\ctxrefines{\emptyset}{e_1}{e_2}{\type}{}$, we mean that the
(limit of the) probability of termination of $\fillctx \ctx [\expr_{1}]$
is below the (limit of the) probability of termination of $\fillctx \ctx [\expr_{2}]$.
The precise definition can be found in \cref{sec:contextual-refinement}; here it suffices to
know that the operational semantics is defined using (probabilistic stateful) schedulers, which
at every step of the execution decide which thread to execute next, and that the termination probability is
obtained by taking the supremum over all possible schedulers (following the notion of \emph{may-termination} in previous work~\cite{nondeterminismLR,prob-nondeterminismLR}).

The central new notion in Foxtrot to reason about contextual refinement is
a \emph{logical concurrent probabilistic refinement judgment}.
The refinement judgment $\refines{\Delta}{\expr}{\expr'}{\type}$
intuitively means that $\expr$ refines $\expr'$ at the type $\type$.
The context $\Delta$ assigns a relational interpretation to all
type variables in $\type$, but can be ignored for now.
The logical refinement provides a sound method to prove contextual refinement,
i.e. $\refines{\emptyset}{e_1}{e_2}{\type}{}$ implies $\ctxrefines{\emptyset}{e_1}{e_2}{\type}{}$.
Following the logical approach to relational refinement \cite{jacm-iris-logrel},
the refinement judgment is defined in the \theaplog logic, which
is in turn defined on top of the Iris base logic \cite{irisjournal}.
The Foxtrot program logic is phrased using Hoare triples
$ \hoare{\prop}{\expr}{\propB}$, which are defined (using ghost state) with an eye for
relational refinement, much as in earlier work \cite{clutch,approxis}.
Rules in \theaplog are often expressed in terms of Hoare triples, so to prove
a refinement judgment, we often unfold its definition to the Hoare triple layer.
We  later see how the refinement judgment is defined with Hoare triples in \cref{sec:logical-relations}.

Roughly speaking, Foxtrot inherits proof rules for concurrency (including
\emph{invariants} and \emph{ghost resources}) from earlier relational logics for concurrency \cite{reloc}, and
 \emph{coupling-based proof rules} for probabilistic reasoning from
earlier relational logics for probability \cite{clutch,approxis}. In fact, these basic
rules are sufficient to verify a wide range of concurrent probabilistic programs, especially
those where the concurrency and probability features do not interact with each other in
complex ways.
However, some programs feature complex behaviors arising from the
complicated interaction between probability and concurrency, e.g.~the $\fimpl$ function from \cref{sec:introduction} where one has to reason about the joint probability distribution produced by sampling in two separate threads.
To address these issues, Foxtrot includes more advanced probabilistic reasoning principles, including
asynchronous couplings using so-called \emph{presampling tapes} (first introduced in
the Clutch logic \cite{clutch}) and \emph{fragmented couplings} for reasoning about
rejection samplers \cite{approxis}. Moreover, inspired by Eris~\cite{eris} and Approxis
\cite{approxis}, Foxtrot also supports 
induction by error amplification through the use of \emph{error credit} resources. We
explain these reasoning principles in \cref{sec:logic}, and we demonstrate their expressiveness in the concurrent setting by using them to reason about a
range of challenging program refinements in \cref{sec:case-studies}, e.g.~an adversarial variant of the von Neumann coin program and the implementation of the \textlog{randombytes\_uniform} function in  the open-source  cryptography software library Sodium. These examples (even including the smaller examples in \cref{sec:logic}) are, to the best of our knowledge, beyond the scope of previous techniques, as they require complex reasoning principles, e.g.~reasoning about higher-order functions, local state, and execution of unknown code. 

In light of this rough overview, the reader may wonder if Foxtrot is a trivial
combination of earlier logics and/or what, if anything, is challenging in its
development? The key technical novelty and
challenges lie in the \emph{definition} of the program logic, specifically in
the weakest precondition predicate used in the definition of Hoare triples, and
the construction of the model of this logic, which ultimately allows us to prove
the soundness theorem above. One of the key challenges is that intuitively, to
show that $\expr_{1}$ refines $\expr_{2}$, for every possible scheduler for the
left-hand side program, we have to exhibit a scheduler for the right-hand side
program $\expr_{2}$, so that $\expr_{1}$ can be coupled with
$\expr_{2}$, and this is done by ``piecing together'' schedulers for
subexpressions (to validate compositional program logic
rules).  This is significantly different from earlier work on
relational logics for concurrency without probabilities~\cite{reloc,
  turon-caresl}, where one only has to construct a single \emph{trace} for the
right-hand side. Now, with probabilistic features, the challenge is that
we have to compose \emph{families} of schedulers (because of the probabilistic
branching we get a family of schedulers). To do so, we end up having to use a
version of the \emph{axiom of choice} in the Iris logic which, to the best of
our knowledge, has not been used in earlier work on Iris-based program logics.
Indeed, \theaplog's model is significantly more sophisticated than that of its
predecessors, such as Clutch~\cite{clutch} and Approxis~\cite{approxis}.
Luckily, users of the logic need not understand the model; it suffices to know
that we have used the model to prove adequacy of the logic (\cref{thm:adequacy}), which intuitively states that proving a Hoare triple in the
\theaplog logic implies that the termination probability of a program is
upper-bounded by that of another, and which is used to prove the soundness
theorem above (\cref{thm:soundness}).

We believe it is a \emph{strong feature} of
Foxtrot that, with the new definition of weakest precondition and model, we are
in fact able to prove soundness of the many strong reasoning principles
mentioned above. We note that this is not automatic (all
the proof rules of Foxtrot have been proved sound from scratch) and not trivial;
in fact, as we later emphasize  in \cref{sec:logic}, one of the proof rules  that one might
expect to hold given earlier work \cite{clutch, approxis}, namely ``presampling on the
right'', is \emph{not} sound in the concurrent setting of Foxtrot.

\paragraph{Contributions} In summary, we provide:
\begin{itemize}
\item The \emph{first} higher-order separation logic, \theaplog, for reasoning about contextual refinement of concurrent probabilistic programs.
\item A rich set of coupling logic rules for advanced probabilistic reasoning, proven sound using the novel model of \theaplog.
\item A collection of case studies showcasing the expressiveness of \theaplog, which are beyond the scope of all previous techniques.
\item All results in the paper are mechanized in the Rocq proof assistant~\cite{coq}, using the Iris separation logic framework~\cite{iris, iris2, iris3, irisjournal} and Coquelicot real analysis library~\cite{coquelicot}.
\end{itemize}

\paragraph{Outline}
We begin by presenting the preliminaries for the rest of the paper (\cref{sec:preliminaries}). We then detail the features of \theaplog and demonstrate the use of these features through small examples (\cref{sec:logic}). Next, we show how \theaplog can scale to verify larger and more involved case studies (\cref{sec:case-studies}). Afterwards, we explain how the semantic model of \theaplog is constructed and proven sound (\cref{sec:model}). We lastly conclude with a discussion of related and future work (\cref{sec:related-work}, \cref{sec:conclusion}).

\section{Preliminaries}
\label{sec:preliminaries}

In this section, we first recall useful facts from probability theory in \cref{sec:prelim-prob}. We then describe the syntax and semantics of our programming language $\thelang$ in \cref{sec:the-lang} and \cref{sec:operational-semantics}. Lastly in \cref{sec:contextual-refinement}, we present the definition of contextual refinement we use in the concurrent probabilistic setting.
\subsection{Probability Theory}
\label{sec:prelim-prob}

A \defemph{subdistribution} (henceforth simply \defemph{distribution}) on a countable set $A$ is a function $\distr : A \ra [0,1]$ s.t.~$\sum_{a \in A}\distr(a) \leq 1$. We write the collection of distributions on $A$ as $\Distr A$. 
Distributions admit monadic operations, with $(\mret\colon A \to \Distr A)$ defined as $\mret(a)(a') \eqdef{} \If{a = a'}then 1 \Else{0}$
and $(\mbind\colon (A \to \Distr B) \times \Distr A \to \Distr B)$ defined as $\mbind(f,\distr)(b) \eqdef{} \sum_{a \in A} \distr(a) \cdot f(a)(b)$.
We often use the notation $\distr \mbindi f$ for $\mbind(f, \distr)$.

Some examples of distributions include the \defemph{null distribution} \(\nulldistr : \Distr A\), given by \(\lambda x . 0\), and the uniform distribution $\unifd{N} \colon \Distr \nat$, which returns $1/(N+1)$ for every $n \in \{0,\dots,N\}$ and $0$ otherwise.

The \emph{expected value} of $X \colon A \to [0,1]$ \wrt $\distr$ is defined as $\expect[\distr]{X} \eqdef{} \sum_{a \in A} \distr(a) \cdot X(a) $.
The \defemph{mass} of \(\distr\) is defined as  $|\distr|\eqdef{}\expect[\distr]{\lambda x. 1}$.%

\theaplog uses \emph{approximate} probabilistic coupling~\cite{Sato:Approximate:2016, approxis} to relate pairs of probabilistic programs.
\begin{definition}[Approximate Coupling]
  Let $\distr_1 \in \Distr{A}$ and $\distr_2 \in \Distr{B}$.
  Given $\err \in [0,1]$ and a relation $R \subseteq A \times B$, we say that there exists an $(\err, R)$-coupling of $\distr_1$ and $\distr_2$ if for all $[0,1]$-valued random variables $X : A \ra [0,1]$ and $Y : B \ra [0,1]$, s.t.~$(a,b) \in R$ implies $X(a)\leq Y(b)$, we have the inequality $ \expect[\distr_1]{X} \leq \expect[\distr_2]{Y} + \err$.
  We write \(\ARcoupl {\distr_1} {\distr_2} \err R\) if such an $(\err,R)$-coupling exists.
\end{definition}

\subsection{The \thelang Language}
\label{sec:the-lang}

Our examples are written in the \thelang language, a higher-order ML-style programming language with higher-order local state, extended with  discrete probabilistic sampling and fork-based unstructured concurrency. The syntax of the language is defined by the grammar below:
\smalldisplay{
\begin{align*}
  \val, \valB \in \Val \bnfdef{}
  & z \in \integer \ALT
  b \in \bool \ALT
  \TT \ALT
  \loc \in \Loc \ALT
  \Rec f \lvar = \expr \ALT
  (\val,\valB) \ALT
  \Inl \val  \ALT
  \Inr \val \ALT
  \\
  \expr \in \Expr \bnfdef{}  &
  \val \ALT
  \lvar \ALT
  \expr_1~\expr_2 \ALT
  \expr_1 + \expr_2 \ALT
  \expr_1 - \expr_2 \ALT
  \ldots \ALT
  \If \expr then \expr_1 \Else \expr_2 \ALT
  (\expr_1,\expr_2) \ALT
  \Fst \expr \ALT \ldots \\
  & \Alloc~\expr \ALT
  \deref \expr \ALT
  \expr_1 \gets \expr_2 \ALT
  \Rand \expr \ALT
  \Fork \expr \ALT
  \Faa \expr_1~\expr_2 \ALT \ldots \\
  \Heap\bnfdef{} & \Loc \fpfn \Val \hspace{4em} \ghostcode{\Tapes\bnfdef{}\Lbl \fpfn \Tape}
  \\
  \state \in \State \bnfdef{} & \Heap \ghostcode{\spac\times\spac\Tapes} \hspace{3em}
  \cfg \in \Conf \bnfdef{}  \List (\Expr) \times \State\\
  \type \in \Type \bnfdef{}
  & \alpha \ALT
    \tunit \ALT
    \tbool \ALT
    \tnat \ALT
    \tint \ALT
    \type \times \type \ALT
    \type + \type \ALT
    \type \to \type \ALT
    \All \alpha . \type \ALT
    \Exists \alpha . \type \ALT
    \tmu \alpha . \type \ALT
    \tref{\type} %
\end{align*}}%
The syntax is mostly standard, e.g.~$\Alloc \expr$, $\deref\expr$, and $\expr_1 \gets \expr_2$ respectively allocate, read from, and write onto a reference.
The expression $\Rand \tapebound$ samples from $\unifd{\tapebound}$,
and $\Fork \expr$ spawns a new thread with $\expr$. \thelang provides various  \emph{atomic} operations for synchronizing threads, e.g.~fetch-and-add $\Faa\expr_1~\expr_2$ adds the integer $\expr_2$ to the value \(\val\) stored at location $\expr_1$ and returns~\(\val\).%

The state $\state$ is given by a pair of maps. The first map models the heap as a finite map from locations to values. The second map is used for keeping track of \emph{presampling tapes}, the discussion of which is postponed until \cref{sec:advanced-rules-foxtrot}.
A program configuration $\cfg \in \List (\Expr) \times \State$ is given by a pair containing the list of currently executing threads and the state.
We say a configuration is \defemph{final} if the first expression in the thread list is a value.

To define contextual refinement of \thelang programs (\cref{sec:contextual-refinement}), we use its typing system, which follows standard ML-style rules.
Typing is expressed as $\pfctx \mid \vctx \proves \expr : \type$ where $\vctx$ is a context assigning types to program variables, and $\pfctx$ contains type variables that may occur in $\vctx$ and $\type$.

\subsection{Operational Semantics}
\label{sec:operational-semantics}

The operational semantics of \thelang programs is defined in 3 steps: (1) a $\stepdistr$ function for stepping a single expression, (2) a stepping function for a configuration \wrt a scheduler, and (3) full program execution \wrt a scheduler\footnote{\thelang is the same language studied in Coneris~\cite{coneris} and we refer readers to that paper for more details.}.

\paragraph{Expressions} We consider a mostly standard call-by-value semantics. The \(\stepdistr \colon (\Expr, \State)\to \DDistr {(\Expr,\State, \List(\Expr))}\) function takes an expression  and a state, and returns a distribution over a new expression, state, and a (possibly empty) list of newly spawned threads.
We display some reductions below; in particular \(\Rand\tapebound\) steps to an integer between $0$ and $\tapebound$ uniformly at random.
\smalldisplay{\begin{align*}
  \stepdistr (\If \True then \expr_1 \Else \expr_2,\state ) &= \mret (\expr_1, \state, [])\\ %
  \stepdistr (\Fork \expr,\state ) &= \mret (\TT, \state, [\expr])\\
  \stepdistr (\Rand \tapebound, \sigma) &= \Lam {(n, \state, [])}\,.\, \textstyle\frac 1 {\tapebound+1} \text{ if } n \in \{ 0, \ldots, N \} \text{\quad and\; \(0\)\; otherwise}
\end{align*}}%

\paragraph{Thread Pools and Schedulers}
The $\tpstepdistr$ function takes a configuration $\cfg=(\vec\expr,\state)$ and an index $\threadid$ and returns a distribution over new configurations after stepping the $\threadid$-th thread:
\smalldisplay{\[
  \tpstepdistr(\vec \expr, \state)(\threadid) \eqdef
  \begin{cases}
    \nulldistr & \text{if \((\vec \expr, \state)\) is final,}\\
    \mret (\vec \expr, \state) & \text{if \(\expr_\threadid \in \Val\) or $j \geq |\vec \expr|$,}\\
    \stepdistr(\expr_\threadid, \state) \mbindi
    \Lam(\expr_\threadid', \state', \vec \expr').
    \mret (\vec \expr [\threadid \mapsto \expr_\threadid'] \dplus \vec \expr' , \state') & \text{otherwise.}
  \end{cases}
\]}%
If \(\cfg\) is final, we return immediately. If \(\expr_\threadid\) is a value or does not exist, we take a stutter step. Otherwise, we step the \(\threadid\)-th thread and add the spawned threads to the thread pool.

The choice of which thread to step is determined by a \emph{scheduler}. A (probabilistic, stateful) scheduler is defined by a transition function $\sch : (\Schstate \times \Cfg) \to \Distr{\Schstate \times \nat}$, which takes in an internal state \(\schstate \in \Schstate\) and a configuration~\(\cfg\), and returns a distribution on its new internal state and the index of the thread to step next. \diffred{Intuitively, the scheduler is used to resolve all non-determinism in a program (which is here limited to which thread is to be executed at each step), and its internal state can be used to fully store the history of the program execution and all the previous non-deterministic results. Here, we consider probabilistic schedulers as it is a fairly standard model in literature (see~\cite{pcol}) and it better reflects some real-world schedulers that make use of randomness in an essential way, e.g.~work-stealing schedulers. }

Given a configuration $\cfg$, a scheduler $\sch$, and a scheduler state $\schstate$, we can now define the single scheduler-step reduction function $\schstepdistr{\sch}{\schstate,\cfg} \in \Distr{\Schstate\times\Cfg}$ as follows:
\smalldisplay{\begin{align*}
  \schstepdistr{\sch}{\schstate,\cfg} \;\eqdef\quad
  \sch (\schstate, \cfg)
  \mbindi  \Lam (\schstate', \threadid) . \tpstepdistr(\cfg, \threadid)
  \mbindi \Lam \cfg' . \mret (\schstate', \cfg')
\end{align*}}%
Informally, $\schstepdistr{\sch}{\schstate,\cfg}$ first evaluates $\sch (\schstate, \cfg)$ to get a new state $\schstate'$ and index~$\threadid$, steps the \(\threadid\)-th thread to obtain the new configuration \(\cfg'\), and returns the new scheduler state and configuration.

\paragraph{Full Program Execution}
We now define \(n\)-step program execution \wrt a scheduler~\(\sch\) with the recursive function  $\execVal_{\sch, n}:(\Schstate\times\Cfg) \to \Distr\Val $
defined below.
Intuitively, $\execVal_{\sch, n}(\schstate, \cfg)(\val)$ represents the probability of returning a value $\val$ in the first thread after at most~$n$ steps of~$\cfg$ \wrt the scheduler~$\sch$ with starting scheduler state~$\schstate$.
The full program execution is then defined as the limit of \(\execVal_{\sch, n}\), which exists by monotonicity and continuity, i.e.~$\limexecVal_{\sch}(\schstate, \cfg) \eqdef \lim_{n\rightarrow \infty} \execVal_{\sch, n} (\schstate, \cfg)$.
\smalldisplay{\[ \execVal_{\sch, n}(\schstate, \cfg) \eqdef
  \begin{cases}
    \mret\val
    & \text{if  \(\cfg = (\val \cons \vec e, \state)\) for some \(\val \in \Val\)}, \\
    \nulldistr
    &  \text{if \(n=0\) and } \cfg \text{ is not final}, \\
    \schstepdistr{\sch}{ \schstate, \cfg} \mbindi \execVal_{\sch, n-1}
    & \text{otherwise.}
  \end{cases} \]}%
The probability that a starting configuration $\cfg$ terminates under scheduler $\sch$ with starting scheduler state $\schstate$ is hence defined as $\schexecTerm_{\sch}(\schstate, \cfg)\eqdef{}  |\limexecVal_{\sch}(\schstate, \cfg)| $.

The upper termination probability of an arbitrary configuration $\cfg$ is taken as the supremum of the termination probability ranging over all possible schedulers and initial scheduler states, i.e.~$\lubexecTerm(\cfg)\eqdef \sup_{\sch, \schstate} \schexecTerm_{\sch}(\schstate,\cfg)$.
We use the upper termination probability $\lubexecTerm$ as the observation predicate for contextual refinement. In other words, we consider \emph{may-termination}~\cite{nondeterminismLR,prob-nondeterminismLR} as the main property of interest for the refinement\footnote{We leave the problem of \emph{must-termination} refinement of concurrent probabilistic programs for future work; we believe an Iris-based logic would appear to require transfinite step-indexing~\cite{transfiris, prob-nondeterminismLR}.}.
\diffred{Defining observational equivalence in terms of (non-)termination is common for non-probabilistic sequential languages, where it is provably equivalent to alternate definitions stated in terms of returning the same Boolean value~\cite{pitts-contextual}. }

We here also provide an auxiliary definition of partial program execution $\pexec_{\sch,n}: (\Schstate\times\Cfg)\rightarrow\DDistr(\Schstate\times\Cfg)$, used in the model of \theaplog.
\smalldisplay{\begin{align*}
  \pexec_{\sch, n}(\schstate, \cfg) &=
                                      \begin{cases}
                                        \mret(\schstate,\cfg) & \text{if \(\cfg\) is final or \(n=0\)}, \\
                                        \schstepdistr{\sch}{ \schstate, \cfg} \mbindi \pexec_{\sch, n-1} & \text{otherwise.}
                                      \end{cases}
\end{align*}}%
We can view \(\pexec\) as a relaxation of \(\execVal\) which keeps probability mass on configurations that are not final, whereas the latter only considers final configurations.

\subsection{Contextual Refinement}
\label{sec:contextual-refinement}

Since we are in a typed setting, we consider only typed contexts.
A program context is well-typed, written $\ctx : (\pfctx \mid \vctx \proves \type) \Ra (\pfctx' \mid \vctx' \proves \type')$, if for any term $\expr$ where $\pfctx \mid \vctx \proves \expr : \type$ we have $\pfctx' \mid \vctx' \proves \fillctx\ctx[\expr] : \type'$.
We say expression $\expr_{1}$ \emph{contextually refines} expression $\expr_{2}$ if for all well-typed program contexts $\ctx$ resulting in a closed program, the upper termination probability of $\fillctx\ctx[\expr_{1}]$ is bounded by the upper termination probability of $\fillctx\ctx[\expr_{2}]$:
\smalldisplay{
  \begin{align*}
  \ctxrefines{\pfctx \mid \vctx}{\expr_{1}}{\expr_{2}}{\type} \eqdef{}
  &\All \type', (\ctx : (\pfctx \mid \vctx \proves \type) \Ra (\emptyset \mid \emptyset \proves \type')), \state . \\
  & \ \lubexecTerm([\fillctx \ctx [\expr_{1}]], \state) \leq \lubexecTerm([\fillctx \ctx [\expr_{2}]], \state)
\end{align*}}%
Note that the statement itself is in the meta-logic (\eg{}, Rocq) and makes no mention of \theaplog{} or Iris.
Contextual equivalence $\ctxeq{\pfctx \mid \vctx}{\expr_{1}}{\expr_{2}}{\type}$ is defined as a refinement in both directions, \ie $(\ctxrefines{\pfctx \mid \vctx}{\expr_{1}}{\expr_{2}}{\type}) \land (\ctxrefines{\pfctx \mid \vctx}{\expr_{2}}{\expr_{1}}{\type})$.
We simply write $\expr_1 \ctxrefinessym \expr_2$ or $\expr_1\ctxeqrel \expr_2$ if the type $\type$ can be inferred  and that $\Gamma$ and $\pfctx$ are empty.

\section{Logic}
\label{sec:logic}

Now we introduce the rules of the \theaplog logic.  We first present a small set of its propositions:
\smalldisplay{\begin{align*}
  \prop,\propB \in \iProp \bnfdef{}
  & \TRUE \ALT \FALSE \ALT \prop \land \propB \ALT \prop \lor \propB \ALT \prop \Ra \propB \ALT
  \All \var . \prop \ALT \Exists \var . \prop \ALT \\
  & \prop \sep \propB \ALT\prop \wand \propB \ALT
  \later \prop \ALT \knowInv{\iname}{\prop} \ALT
  \ownGhost{\gname}{\ghostRes} \ALT\pvs[\mask_1][\mask_2] \prop \ALT \tid\tpmapsto \expr \ALT
  \loc\mapsto\val \ALT \loc\specmapsto \val \ALT\\
  &
    \hoare{\prop}{e}{\propB}[\mask] \ALT \pupd\prop\ALT  \progtape{\lbl}{\tapebound}{\tape} \ALT \spectape{\lbl}{\tapebound}{\tape} \ALT \upto{\err} \ALT
    \ldots
\end{align*}}%
\theaplog is built on top of the Iris base logic~\cite{iris} and it inherits many of the basic propositions found in standard Iris separation logics, such as the \emph{later} modality $\later$, \emph{invariants} $\knowInv{\iname}{I}$, \emph{ghost resources} $\ownGhost{\gname}{\ghostRes}$, and \emph{fancy updates} $\pvs[\mask_1][\mask_2]\prop$. 
The specification program on the right-hand side is represented by the specification thread connective $\tid\tpmapsto \expr$, which intuitively states that the $\tid$-th right-hand side thread is currently executing expression $\expr$. We use the points-to connectives $\loc\mapsto\val$ and $\loc\specmapsto \val$ to represent exclusive ownership of the location $\loc$ storing value $\val$ in the left-hand side program and right-hand side program, respectively. 

Central to \theaplog's logic is the Hoare triple proposition. These Hoare triples are annotated  with a mask $\mask$ for tracking opened invariants \diffred{(we omit it if it is the top mask $\top$)}. We give meaning to the \theaplog Hoare triple via the adequacy theorem \cref{thm:adequacy}:
\begin{theorem}[Adequacy]\label{thm:adequacy}
  If we can prove $\hoare{%
    0\tpmapsto \expr' } {\expr}{\val. \Exists \val'. 0\tpmapsto \val'}$ in \theaplog,
  then for all states $\state, \state'$, we have $\lubexecTerm([\expr], \state)\leq\lubexecTerm([\expr'], \state')%
  $.
\end{theorem}
\diffred{Roughly speaking, the theorem states that if we are able to prove the Hoare triple for the expression $\expr$ assuming that the $0$-th thread on the right stores $\expr'$ initially, and we can match up the execution traces of $\expr$ and $\expr'$ such that whenever we have a trace where $\expr$ terminates, the corresponding trace for $\expr'$ also terminates, then the upper termination probability of $\expr$ is upper-bounded by that of $\expr'$.}

\theaplog also supports other propositions for advanced probabilistic reasoning such as the \emph{probabilistic update modality} $\pupd\prop$, \emph{presampling-tape} connectives $\progtape{\lbl}{\tapebound}{\tape}$ and $\spectape{\lbl}{\tapebound}{\tape}$, as well as \emph{error credits} $\upto{\err}$. We describe these in more detail later in this section.

In \cref{sec:logical-relations}, we first present the logical relation that gives rise to the refinement judgment introduced in \cref{sec:key-ideas}, as well as the key results of the refinement judgment. We then present an overview of the rules of \theaplog, from basic program-logic rules in \cref{sec:basic-rules-foxtrot} to more advanced rules for more complicated couplings in \cref{sec:advanced-rules-foxtrot}. We emphasize that though various rules and features are inspired by  prior work, as mentioned in \cref{sec:key-ideas}, proving all the rules are sound in our novel \theaplog model is highly \emph{non-trivial} and is \emph{not automatic}, which we explain more in detail in \cref{sec:model}.

\subsection{Logical Relations}
\label{sec:logical-relations}
In \cref{sec:key-ideas}, we introduced the logical concurrent probabilistic refinement judgment for proving contextual refinement of \thelang programs. In \theaplog, this judgment  $\refines{\Delta}{\expr}{\expr'}{\type}$ is defined as $ \All \lctx,\ \tid. \hoare{\tid\tpmapsto\fillctx \lctx [\expr']} {\expr}{\val.\Exists \val'. \tid\tpmapsto\fillctx \lctx [\val'] \sep \semInterp{\type}{\Delta}{\val,\val'}}$, where $\lctx$ quantifies over all evaluation contexts. Here $\semInterpS{\type}{\Delta}$ represents the semantic interpretation of type $\type$ defined inductively on the type, which intuitively describes when two values of a certain type are considered related. The way we define the logical relations and encode it within the logic is standard and similar to that in previous work~\cite{jacm-iris-logrel}. 
We then extend the refinement judgment to open terms in the standard way, \ie
$ \refines{\Delta \mid \Gamma}{\expr}{\expr'}{\type} \eqdef{} \forall \gamma, \gamma'. \semInterp{\Gamma}{\Delta}{\gamma,\gamma'} \wand \refines{\Delta}{\expr\gamma}{\expr'\gamma'}{\type}$, and prove the fundamental lemma of the binary logical relations:

\begin{lemma}[Fundamental Lemma of Binary Logical Relations]\label[lemma]{thm:fundamental-binary}
  If\; \(\pfctx\mid\Gamma \vdash \expr : \tau\)\; then for all $\Delta$ assigning a relational interpretation to every type variable in $\pfctx$, we have \(\refines {\Delta\mid\Gamma} \expr \expr \tau\)\,.
\end{lemma}
We then show that the refinement relation is sound with respect to contextual refinement: 

\begin{theorem}[Soundness]
  \label{thm:soundness}
  Let $\pfctx$ be a type variable context, and assume that, for all $\Delta$ assigning a relational interpretation to all type variables in $\pfctx$, we can derive $\refines{\Delta\mid\vctx}{e_1}{e_2}{\type}{}$.
  Then $\ctxrefines{\pfctx\mid\vctx}{e_1}{e_2}{\type}{}$.
\end{theorem}

\diffred{Here, note that the type system of \thelang is considerably rich, supporting complex features such as polymorphism and higher-order references.
  Therefore, later in the paper, even though the statement of our examples might not directly involve these features, the contexts that interact with them might.
Hence, to prove contextual refinement, our  logic \theaplog has to be, and is, expressive enough to support higher-order propositions and define our logical relations (which utilizes second-order impredicative quantification).}

In addition to the binary refinement relations, we also additionally construct a \emph{unary} logical relations model for proving type safety of \thelang programs where $\relates{\Delta}{\expr}{\type}\eqdef  \hoare{\TRUE}{\expr}{\val. \semInterp{\type}{\Delta}{\val}}$. This is then similarly extended to open terms: $\relates{\Delta\mid\Gamma}{\expr}{\type}\eqdef \forall \gamma. \semInterp{\Gamma}{\Delta}{\gamma} \wand \relates{\Delta}{\expr\gamma}{\type} $ and we prove its expected fundamental lemma, which is used later in the adversarial von Neumann coin example in \cref{sec:vncoin} to reason about unknown adversaries.
\begin{lemma}[Fundamental Lemma of Unary Logical Relations]\label[lemma]{thm:fundamental-unary}
  If\; \(\pfctx\mid\Gamma \vdash \expr : \tau\)\; then for all $\Delta$ assigning a unary interpretation to all type variables in $\pfctx$, we have \(\relates {\Delta\mid\Gamma} \expr  \tau\)\,.
\end{lemma}

\subsection{Basic Rules}
\label{sec:basic-rules-foxtrot}
In this subsection, we cover  structural rules, rules for symbolically executing the left or right-hand side programs, and the standard rule for coupling programs.
\paragraph{Standard Iris Rules}
\theaplog inherits many standard rules in similar Iris separation logics, such as \ruleref{ht-bind} for reasoning about sequences of instructions and the frame rule \ruleref{ht-frame} for framing resources around a Hoare triple. We symbolically execute the left-hand side program with familiar-looking rules such as that of \ruleref{ht-load} and \ruleref{ht-fork}. As mentioned in \cref{sec:key-ideas}, \theaplog is a concurrent separation logic and it supports  invariants (with rules such as  \ruleref{ht-inv-alloc} and \ruleref{ht-inv-open}) and ghost resources (standard rules omitted here for brevity).
\smalldisplay{
\begin{mathpar}
  \inferH{ht-bind}
  { \hoare{\prop}{\expr}{\val . \propB} \\  \All \val. \hoare{\propB}{\fillctx\lctx[\val]}{\propC}}
  { \hoare{\prop}{\fillctx\lctx[\expr]}{\propC}}
  \and
  \inferH{ht-frame}
  {\hoare{\prop}{\expr}{\propB}}
  { \hoare{\prop\sep\propC}{\expr}{\propB\sep\propC}}
  \and
  \inferH{ht-load}
  {}
  {\hoare{l\mapsto v}{\deref l}{w.w=v\sep l\mapsto v}}
  \and
  \inferH{ht-fork}
  {\hoare{\TRUE}{\expr}{\TRUE}}
  {\hoare{\TRUE}{\Fork \expr}{w.w=\TT}}\and
  \inferH{ht-inv-alloc}
  { %
    \hoare{\knowInv{\iname}{\prop} \sep \propB}{\expr}{\propC} }
  { \hoare{\prop \sep \propB}{\expr}{\propC} }
  \and
  \inferH{ht-inv-open}
  {\expr\spac \atomic \\
    \hoare{\later I\sep\prop}{\expr}{\later I\sep\propB}[\mask]}
  { \hoare{\knowInv{\iname}{I} \sep \prop}{\expr}{\propB}[\mask\uplus\{\iname\}]}
\end{mathpar}}%
We  implement the parallel composition operator with the $\Fork$ primitive, and with ghost resources, we prove the  familiar-looking rule \ruleref{ht-par-comp} for parallel composition: 
\smalldisplay{\[
  \infrule[right]{ht-par-comp}
          { \hoare{\prop_1}{\expr_1}{\val_1 \ldotp \propB_1\  \val_1 } \\
             \hoare{\prop_2}{\expr_2}{\val_2 \ldotp \propB_2\  \val_2 }
          }
  { \hoare{\prop_1\sep \prop_2}{\Parallel{\expr_1}{\expr_2}}{(\val_1, \val_2) \ldotp \propB_1\  \val_1\sep \propB_2\ \val_2 } }
  \]}%
\paragraph{Probabilistic Update Modality}
We previously saw that we can symbolically execute the left-hand side program with the Hoare triple directly. Naturally, one might ask,  how can we do so for the right? The answer is that we do so with the \emph{probabilistic update modality} $\pupd$, a construct first introduced in Coneris~\cite{coneris} for reasoning about randomized logical atomicity.  Here, we extend the modality  for  manipulating the right-hand side program; we also later see how the probabilistic update modality is used in more complicated rules, such as those for presampling onto tapes and for generating error credits. 

The probabilistic update modality supports monadic-like rules, i.e.~\ruleref{pupd-ret}  and \ruleref{pupd-bind}. Most importantly, the probabilistic update modality can be eliminated in the precondition of a Hoare triple (\ruleref{ht-pupd-elim}). As a consequence,  rules for symbolically executing the right-hand side program can be succinctly expressed with the probabilistic update modality, which is  different in presentation from earlier relational logics~\cite{reloc, jacm-iris-logrel}.
For example, \ruleref{pupd-load} and \ruleref{pupd-fork} symbolically steps a load and $\Fork$ expression on the right-hand side program, respectively.
\smalldisplay{\begin{mathpar}
	\infrule[right]{pupd-ret}
  { \prop
  }
  {\pupd\prop}
  \and
	\infrule[right]{pupd-bind}
  { \pupd\prop \\
    \prop \wand \pupd \propB
  }
  {\pupd \propB}
  \and
	\infrule[right]{ht-pupd-elim}
  {\hoare{\prop \sep \propB}{\expr}{\propC}
  }
  {\hoare{(\pupd\prop) \sep \propB}{\expr}{\propC}}
  \\
	\infrule[right]{pupd-load}
         {l\spointsto v \\
           \tid\tpmapsto \fillctx \lctx [\deref l]}
         {\pupd{(l\spointsto v \sep \tid\tpmapsto \fillctx \lctx [v])}}
         \and
	\infrule[right]{pupd-fork}
         {\tid\tpmapsto \fillctx\lctx [\Fork \expr]}
         {\pupd{(\tid\tpmapsto\fillctx\lctx[\TT]\sep \Exists \tid'. \tid'\tpmapsto \expr)}}
\end{mathpar}}%
In addition, \theaplog provides  \ruleref{pupd-par-comp} for parallel composition on the specification side:
\smalldisplay{\[
	\infrule[right]{pupd-par-comp}
       {\tid\tpmapsto\fillctx\lctx[\Parallel{\expr_1}{\expr_2}]}
       {\pupd{
           \Exists \tid_1, \tid_2, \lctx_1, \lctx_2.
           \left(\begin{array}{l}
             \tid_1\tpmapsto\fillctx\lctx_1[\expr_1] \sep 
             \tid_2\tpmapsto\fillctx\lctx_2[\expr_2] \sep \\
             \All \val_1, \val_2.
             \tid_1\tpmapsto\fillctx\lctx_1[\val_1] \wand
             \tid_2\tpmapsto\fillctx\lctx_2[\val_2] \wand \pupd{\tid\tpmapsto\fillctx\lctx[(\val_1,\val_2)]}
           \end{array}
           \right)
       }}
       \]}%
\paragraph{Couplings}
\theaplog makes use of (approximate) \emph{probabilistic couplings}~\cite{thorisson/2000, lindvall_lectures_2002} to relate the randomness of two programs. We present the following simple yet useful coupling rule \ruleref{ht-couple}, which
allows us to reason as if the two sampled values on both programs are related by a bijective function  $f$ under the same sampling space $\{0,\dots,\tapebound\}$. 
\smalldisplay{\begin{mathpar}
	\infrule[right]{ht-couple}
         { f~\text{bijection}
           \\
           \All n \leq \tapebound . \hoare{\tid\tpmapsto\fillctx\lctx [f(n)]}{n}{\Phi} }
  { \hoare{\tid\tpmapsto \fillctx \lctx [\Rand\tapebound] }{\Rand \tapebound}{\Phi} }
\end{mathpar}}%
Together with the other logical facilities for reasoning about concurrency, e.g.~invariants,  the coupling rule \ruleref{ht-couple} is sufficient for reasoning about various concurrent probabilistic programs. 
As an example, consider the left-hand side entropy-mixer program in \cref{fig:entropy-mixer}, originally  studied in Probabilistic Concurrent Outcome Logic~\cite{pcol}. Here, we are given two sources of entropy, $x_1$ and $x_2$, the former being unreliable as it can be adversarially controlled by the scheduler, while $x_2$ is of high quality. Nonetheless, we are still able to ``combine'' these two sources and produce a high-quality output that  behaves like a $\Rand 1$. We can verify this claim by showing contextual equivalence between the entropy-mixer program and a $\Rand 1$ program with \theaplog. Even though this program uses both concurrency (parallel composition) and random sampling ($\Rand$), the bulk of the proof simply boils down to applying the soundness theorem (\cref{thm:soundness}), defining an invariant to share the reference $y$ between the concurrent threads, and applying  \ruleref{ht-couple} with the right bijective function $f$.

\begin{figure}[ht!]
  \small
  \begin{minipage}[c]{0.33\linewidth}
    \begin{align*}
      &\Let (y,r) = (\Alloc 0, \Alloc 0) in \\
      & \biggl(\Parallel{
        \begin{array}{l}
          \Let x_1 =\deref y in \\
          \Let x_2 = \Rand 1 in \\
          r \gets ((x_1+x_2)~\textlog{mod}~2)
          \end{array}
      }{y \gets 1} \biggr); \\
      & \deref r
    \end{align*}
    \end{minipage}
  \begin{minipage}[c]{0.1\linewidth}
    \begin{align*}
      &\ctxeqrel
      \end{align*}
    \end{minipage}
  \begin{minipage}[c]{0.33\linewidth}
    \begin{align*}
      &\Rand 1
    \end{align*}
    \end{minipage}
  
  \caption{Entropy mixer example.\label{fig:entropy-mixer} }
\end{figure}
We summarize this subsection with the following slogan:
\begin{result}
  \begin{center}
\textbf{\hypertarget{slogan1}{Slogan 1}: Invariants, ghost resources,  and \ruleref{ht-couple} are (almost) all you need.}
  \end{center}
\end{result}

\subsection{Advanced Rules }
\label{sec:advanced-rules-foxtrot}
Previously, we saw the standard coupling rule \ruleref{ht-couple} for coupling a $\Rand \tapebound$ on the left with one on the right under a bijective function. However, unsurprisingly, there are times where it is not the case that there is conveniently  a single $\Rand$ of the same range on both programs.  In those cases, \theaplog uses other advanced logical facilities and rules to establish the couplings.
\diffred{In the following examples, we also move away from reasoning about programs of base types (e.g.~$\tnat$), to reasoning about programs of function types (e.g.~$A\ra B$),  which requires nested Hoare triples. }

\newcommand{\batchimpl}{\textlog{batchImpl}}
\newcommand{\batchspec}{\textlog{batchSpec}}
\newcommand{\batchimplt}{\textlog{batchImpl'}} %
\paragraph{Combining Samplings in Separate Threads}
\theaplog supports rules for reasoning about programs that perform random samplings across different threads.
Consider the batch sampling program $\batchimpl$ in \cref{fig:batch-sampling}, which is a generalization of $\fimpl$ from \cref{sec:introduction}. It computes $\Rand \tapebound$ and $\Rand \tapeboundB$ concurrently and returns the sum of the result with the first value multiplied by $(\tapeboundB+1)$. In other words, it aggregates two parallelized random samples to produce another one. We can show that this function returns a number that is uniformly distributed by showing its contextual equivalence to the function $\batchspec$ that directly samples from $\Rand ((\tapebound+1)\cdot(\tapeboundB+1)-1)$.
\begin{figure}[ht!]
  \small
  \begin{minipage}[c]{0.4\linewidth}
    \begin{align*}
      & \batchimpl \eqdef{}\\
      & \spac \Lam \_. \Let (x,y) = (\Parallel{\Rand\tapebound}{\Rand\tapeboundB}) in%
      \ x\cdot(\tapeboundB+1) + y
    \end{align*}
    \end{minipage}
  \begin{minipage}[c]{0.1\linewidth}
    \begin{align*}
      &\ctxeqrel
      \end{align*}
    \end{minipage}
  \begin{minipage}[c]{0.1\linewidth}
    \begin{align*}
      & \batchspec \eqdef{}\\
      &\Lam \_. \Rand  ((\tapebound+1)\cdot(\tapeboundB+1)-1)
    \end{align*}
    \end{minipage}
  
  \caption{Batch-sampling example.}
  \label{fig:batch-sampling}
\end{figure}

Let us consider the left-to-right refinement first. To reason about random samplings occurring in different threads in $\batchimpl$, \theaplog uses presampling tapes (first introduced in Clutch~\cite{clutch}) to ``compress'' multiple samplings and reason about them as if they are one.

\smalldisplay{\begin{minipage}{0.5\linewidth}
  \begin{align*}
  \val \in \Val \bnfdef{}& \ldots \ALT \lbl \in \Lbl \\
  \expr \in \Expr \bnfdef{}& \ldots \ALT
                             \AllocTape\,\expr \ALT
                             \Rand \expr_{1}~\expr_{2}
  \end{align*}
\end{minipage}
\hfill
\begin{minipage}{0.5\linewidth}
  \begin{align*}
  \state \in \State \eqdef{}& (\Loc \fpfn \Val) \times (\Lbl \fpfn \Tape) \\
    t \in \Tape \eqdef{}& \{ (\tapebound, \tape) \mid \tapebound \in \mathbb{N} \wedge \tape \in \mathbb{N}_{\leq \tapebound}^{\ast} \}
  \end{align*}
\end{minipage}}%

As mentioned previously in \cref{sec:the-lang}, the state of our language contains a finite map from tape labels to presampling tapes. We also introduce two new constructs in \thelang, $\AllocTape \expr$ and $\Rand \expr_1~\expr_2$. The expression $\AllocTape \tapebound$ allocates an empty tape and returns a new tape label \eqref{eq:step:alloctape}. In addition to the regular unlabelled random sampling we have shown before, we can perform a \emph{labelled} random sampling by supplying a tape label to the $\Rand$ operation. If the tape label points to an empty tape, then the labelled random sampling instruction acts like a normal unlabelled one \eqref{eq:step:randnil}, \ie it samples a number uniformly from the range. However, if it is not empty, we pop and return the first number in the tape \emph{deterministically} \eqref{eq:step:randcons}. Perhaps surprisingly, there are \emph{no} constructs in \thelang for presampling elements onto tapes, and so, tapes remain empty throughout the execution of the program  (and hence, \eqref{eq:step:randcons} is never applied normally). Instead, the tape presampling action is done only within the logic during the proof.
\smalldisplay{\begin{align}
  \stepdistr(\AllocTape \tapebound, \state) &= \mret (\lbl,\lupdate{\state}{\lbl}{(\tapebound,\nil)} , []) \quad \text{(where $\lbl$ is fresh \wrt \( \state\))} \label{eq:step:alloctape}\\
  \stepdistr (\Rand \lbl\ \tapebound, \sigma) &= \Lam {(n, \state, [])}\,.\, \If  0\leq n \leq N then \textstyle\frac 1 {\tapebound+1} \Else 0 \quad \text{(where $\state[\lbl] = (\tapebound, \nil)$)}\label{eq:step:randnil} \\
  \stepdistr(\Rand \lbl\ \tapebound, \sigma) &= \mret (n,\lupdate{\sigma}{\lbl}{\tape} , [])\quad \text{(where $\state[\lbl]=(\tapebound,  n\lapp\tape)$)} \label{eq:step:randcons}
\end{align}}%
Just like we have the heap points-to connectives on both sides of the refinement, we have the tapes connectives for both sides as well:  $\progtape{\lbl}{\tapebound}{\tape}$ and $\spectape{\lbl}{\tapebound}{\tape}$. The rules \ruleref{ht-alloc-tape} and \ruleref{pupd-alloc-tape} allocates a new empty tape, and \ruleref{ht-rand-tape} returns the first element of a tape deterministically for a labelled $\Rand$  on the left-hand side program. There is however \emph{no} counterpart rule of \ruleref{ht-rand-tape} for the right-hand side program, as presampling onto tapes on the specification side is unsound\footnote{\diffred{Even though presampling onto tapes on the specification side is unsound, the (empty) specification-side tape connective is still needed for \ruleref{ht-couple-rand-lbl}. }}. 
\smalldisplay{\begin{mathpar}
	\inferH{ht-alloc-tape}
  {}
  {\hoare{\TRUE}{\AllocTape~\tapebound}{ \lbl \ldotp \progtape{\lbl}{\tapebound}{\nil}}}
  \and
	\inferH{pupd-alloc-tape}
  {\tid\tpmapsto \fillctx\lctx[\AllocTape~\tapebound]}
  {\pupd { \Exists \lbl .
      \tid\tpmapsto \fillctx\lctx[\lbl]\sep\spectape{\lbl}{\tapebound}{\nil}}}
  \and
	\inferH{ht-rand-tape}
  { }
  { \hoare
    {\progtape{\lbl}{\tapebound}{n\cdot\tape}}
    {\Rand \lbl~\tapebound}
    {\Ret x .  x = n \sep \progtape{\lbl}{\tapebound}{\tape}} }
\end{mathpar}}%
We emphasize  that presampling tapes are only used for verification purposes and they do not change the behavior of programs (recall that a labelled $\Rand$ with an empty tape behaves the same as an unlabelled one). 
To relate a program annotated with tapes and one without, we provide the two rules \ruleref{ht-couple-rand-lbl} and \ruleref{pupd-couple-lbl-rand} to couple a pair of labelled and unlabelled $\Rand$ operations. The rule \ruleref{ht-couple-rand-lbl} couples an unlabelled $\Rand$ on the left and a labelled one on the right with the empty tape resource on the specification side provided. On the other hand, the rule \ruleref{pupd-couple-lbl-rand} couples a presampling action of a tape resource on the implementation side and a $\Rand$ construct on the right; after which, applying \ruleref{ht-rand-tape} later reads out the presampled value in the tape resource. 
\smalldisplay{\begin{mathpar}
  \inferH{ht-couple-rand-lbl}
         {
           \begin{matrix}
           f~\text{bijection}\cr
           \All n \leq \tapebound . \hoare{\tid\tpmapsto\fillctx\lctx [f(n)]\sep\spectape{\lbl}{\tapebound}{\nil}}{n}{\Phi}
           \end{matrix}
         }
	 { \smash{\hoare{\tid\tpmapsto \fillctx \lctx [\Rand\lbl~\tapebound]
           \sep
           \spectape{\lbl}{\tapebound}{\nil}}
	   {\Rand \tapebound}{\Phi} }}
         \quad
  \inferH{pupd-couple-lbl-rand}
         { f~\text{bijection}\;\;
           \tid\tpmapsto \fillctx \lctx [\Rand\tapebound]
           \;\;
           \progtape{\lbl}{\tapebound}{\tape}
           }
	 { \smash{\pupd \Exists n\leq\tapebound. 
           \progtape{\lbl}{\tapebound}{\tape\lapp[n]}\sep
	   \tid\tpmapsto \fillctx \lctx [f(n)]}
         }
\end{mathpar}}%
Returning to our example, we prove the refinement via the following intermediate program $\batchimplt$ annotated with tapes.
 \smalldisplay{   \begin{align*}
      \batchimplt\eqdef{}
      & \Lam \_.  \ghostcode{\Let (\lbl,\lbl')=(\AllocTape\tapebound, \AllocTape\tapeboundB) in} \\
      & \hspace{1.7em}  \Let (x,y) = (\Parallel{\Rand\ghostcode{\lbl}~\tapebound}{\Rand\ghostcode{\lbl'}~\tapeboundB}) in \\
      & \hspace{1.75em}  x\cdot(\tapeboundB+1) + y
    \end{align*}}%
  We split the refinement into two parts by transitivity of contextual refinement.

  \diffred{In the first part, we  show that $\batchimpl$ refines the intermediate program $\batchimplt$. This is relatively straightforward as the structure of the two programs are similar; After allocating the specification tapes with \ruleref{pupd-alloc-tape}, we  apply \ruleref{ht-couple-rand-lbl} to couple the unlabelled $\Rand$s on the left with labelled ones on the right. }

  \diffred{Next, we show that the intermediate program  $\batchimplt$ refines $\batchspec$.
    We first allocate the tapes in $\batchimplt$ with \ruleref{ht-alloc-tape}. We can take advantage of the two tape resources and perform a coupling, where we couple two presampling operations on the left tapes with an unlabelled $\Rand$ on the right with the following rule \ruleref{pupd-couple-two-rands-rand}, choosing $f(n,m)\eqdef{}n\cdot(M+1)+m$. Subsequently, each thread in $\batchimplt$ deterministically reads from their corresponding tape resource with \ruleref{ht-rand-tape}.}
 \smalldisplay{\begin{mathpar} \inferH{pupd-couple-two-rands-rand}
         { \tid\tpmapsto \fillctx \lctx [\Rand((\tapebound+1)\cdot(\tapeboundB+1)-1)]
           \\
           \progtape{\lbl}{\tapebound}{\tape}\\
           \progtape{\lbl'}{\tapeboundB}{\tapeB}\\
           f~\text{bijection}}
         { \pupd\Exists n\leq\tapebound, m\leq\tapeboundB.
           \tid\tpmapsto \fillctx \lctx [f(n,m)]\sep
           \progtape{\lbl}{\tapebound}{\tape\lapp[n]}\sep
           \progtape{\lbl'}{\tapeboundB}{\tapeB\lapp[m]}
         }
 \end{mathpar}}%
 \diffred{At a high level, the key step of this proof is that we take advantage of presampling tapes to perform multiple (possibly future) random samplings in advance, and to couple this operation with a single sampling on the right. }
 
 Now that we proved the left-to-right refinement, one might ask, can we adopt a similar proof structure with tape presampling for the right-to-left refinement? Unfortunately the answer is \emph{no}! It turns out that, unlike in previous logics supporting  tapes~\cite{clutch, approxis}, it is \emph{unsound} to perform any presampling on the right-hand side program in \theaplog, and we provide a proof of this in \appref{app:counter}. \diffred{Intuitively, presampling is an operation that enables us to predict future random outcomes from $\Rand$. Allowing arbitrary presampling operations on the right commutes probabilistic and non-deterministic choices in a program, and as a result, the scheduler constructed by the model is one that can make scheduling decisions based on future probabilistic outcomes, which is not valid according to the operational semantics of \thelang. In previous logics that support presampling on the right, e.g.~Clutch~\cite{clutch} and Approxis~\cite{approxis}, this paradox does not occur since the underlying language is sequential, and the operational semantics of programs does not involve any scheduler to resolve non-deterministic choices (e.g.~thread scheduling). }

 That said, we can still show that $\batchspec$ refines $\batchimpl$ without presampling on the right-hand side program. To see why, consider symbolically stepping through $\batchimpl$ now residing on the right-hand side of our refinement. By applying \ruleref{pupd-par-comp}, we then create two thread resources via the parallel composition operator. We now have access to both $\Rand$ operations as specification resources at the same time, and we can couple the single $\Rand$ operation in $\batchspec$ with the two $\Rand$ operations residing in two \emph{separate} threads on the right with the following rule \ruleref{ht-couple-rand-two-rands}.
\smalldisplay{\begin{mathpar}
  \inferH{ht-couple-rand-two-rands}
         { 
           f~\text{bijection}\\
           \All n \leq \tapebound, m\leq\tapeboundB .
           \hoare{\tid\tpmapsto\fillctx\lctx [n]\sep
           \tid'\tpmapsto\fillctx\lctx' [m]}
           {f (n,m)}{\Phi} }
         { \hoare{\tid\tpmapsto \fillctx \lctx [\Rand\tapebound]
           \sep
           \tid'\tpmapsto \fillctx \lctx' [\Rand\tapeboundB]}{\Rand ((\tapebound+1)\cdot(\tapeboundB+1)-1)}{\Phi} }
     \end{mathpar}}%
\diffred{We note that \ruleref{pupd-couple-two-rands-rand} and \ruleref{ht-couple-rand-two-rands}  can be generalized in a straightforward manner to allow couplings between multiple random samplings on one side with one on the other. This enables  us to show contextual refinement between parallel programs that utilize more than two threads. }

Let us take a step back and summarize the key ideas of this example. The main challenge of this batch sampling example is that of relating two $\Rand$ operations with one. Because there is always one left-hand side expression in a Hoare triple, in order for us to reason about multiple $\Rand$ operations in different threads on the left-hand side program, we utilize presampling tapes and presampling coupling rules to perform the coupling in advance before the actual samplings on the left occur. When the multiple $\Rand$ operations occur in different threads on the right, we can directly make use of the multiple thread resources produced by the concurrent nature of the program to perform the coupling. We summarize these ideas with the following slogan.
\begin{minipage}{\textwidth}
\begin{result}
    \begin{center}

      \textbf{\hypertarget{slogan2}{Slogan 2}: Tape the left; parallelize the right. }
      \end{center}
\end{result}
\end{minipage}
\paragraph{Rejection Sampling}
We now turn our attention to another difficult-to-verify pattern in probabilistic computation: rejection sampling schemes. 
Rejection samplers are a class of Las Vegas algorithms; they generate random samples from a target distribution by repeatedly sampling from another distribution and only accepting a sample if it satisfies certain conditions. Though this rejection sampling technique is ubiquitous, it is also challenging to reason about, particularly due to its use of unbounded looping in the implementation.

Consider the rejection sampler program on the left-hand side of the refinement in \cref{fig:rejection-sampler}. Given $\tapeboundB<\tapebound$, the rejection sampler program is a function that first samples from $\{0,\dots,\tapebound\}$. We then return the value if it is upper bounded by $\tapeboundB$, otherwise we repeat the process. It should not be surprising that this function returns an integer from $\{0,\dots,\tapeboundB\}$ uniformly, and hence one would want to show that it is contextually equivalent to one that calls $\Rand \tapeboundB$. We emphasize that this result is \emph{novel} and is stronger than those found in previous work~\cite{approxis}, as we show that this contextual equivalence holds even in the concurrent setting. 
\begin{figure}[ht!]
  \small
  \begin{minipage}[c]{0.4\linewidth}
    \begin{align*}
      &\Rec f~\_ =  \Let x = \Rand \tapebound in \\
      &\spac \If x \leq \tapeboundB then x \Else f~\TT
    \end{align*}
    \end{minipage}
  \begin{minipage}[c]{0.1\linewidth}
    \begin{align*}
      &\ctxeqrel
      \end{align*}
    \end{minipage}
  \begin{minipage}[c]{0.3\linewidth}
    \begin{align*}
      &\Fun  \_ . \Rand \tapeboundB
    \end{align*}
    \end{minipage}  
  \caption{Rejection sampler example.}
  \label{fig:rejection-sampler}
\end{figure}

Here, we sadly cannot use the trick presented in \hyperlink{slogan2}{\textbf{Slogan 2}}; instead of trying to relate two $\Rand$s with one,  we are trying to relate a pair of $\Rand$s in a somewhat unnatural manner. If the rejection sampler samples a value $\leq\tapeboundB$, we want that value to be coupled with $\Rand\tapeboundB$. Otherwise if we sampled a value $>\tapeboundB$, we do not want to step $\Rand\tapeboundB$ at all. In other words, we only want to perform a coupling with the last sampled value in the rejection sampler that gets accepted with $\Rand\tapeboundB$.

\theaplog solves this issue by supporting \emph{fragmented couplings}, a  form of couplings first introduced in Approxis~\cite{approxis}\footnote{The presentation of the fragmented coupling rules in \theaplog is however slightly different from those in Approxis. This is because the Approxis fragmented coupling rules are expressed with presampling on the right-hand side program, which is unsound in \theaplog as shown in \appref{app:counter}.}. To see fragmented couplings in action, let us first consider the left-to-right direction of the refinement.  After stepping through the program where we reach $\Rand\tapebound$ on the left and $\Rand\tapeboundB$ on the right, we apply \ruleref{ht-couple-fragmented} to couple the $\Rand$ expressions in a fragmented way, taking $f$ as the identity function.
\smalldisplay{\[
  \inferH{ht-couple-fragmented}
         { \tapeboundB\leq\tapebound\\
           f~\text{injective}\\
           \All n \leq \tapebound.
           \hoare{%
             \begin{array}{l}(\Exists m\leq\tapeboundB. f(m)=n\sep\tid\tpmapsto \fillctx \lctx [m]
             )\lor \cr
             (\neg(\Exists m\leq\tapeboundB. f(m)=n) \sep
           \tid\tpmapsto \fillctx \lctx [\Rand\tapeboundB]
           )
           \end{array}
           }{n}{\Phi} }
         { \hoare{\tid\tpmapsto \fillctx \lctx [\Rand\tapeboundB]}{\Rand \tapebound}{\Phi} }
         \]}%
Intuitively, \ruleref{ht-couple-fragmented} performs a case split on the value of $\Rand\tapebound$. If it is accepted, then we also take a step for $\Rand\tapeboundB$ on the right, and their values are related by some injective function $f$ (the identity function for this example). Otherwise, we do not take a step on the right. From here, we continue stepping through the proof after a case split. If we sample an accepted value, we can establish the postcondition directly. In the other case, we step  the left-hand side program and return to the start of the loop, at which point, we can apply the hypothesis of Löb induction, a standard rule for reasoning about recursive functions in Iris logics~\cite{irisjournal}.

Again, one might ask, can we do the same for the other right-to-left direction of the refinement? Alas, reality is often disappointing and the answer is \emph{no}. This is due to two reasons. The first reason is more technical; we can perform a fragmentation step on the right-hand side program with \ruleref{ht-couple-fragmented} where we can choose to step the program or not, but we cannot do something similar for the left program due to how the \theaplog model is defined. Secondly, we cannot apply Löb induction for this direction; in the case that the right-hand side program samples a rejected value, we can only proceed by stepping the right-hand side program back to the start of the recursive function with the left-hand side program kept unchanged. We hence cannot eliminate the later modality in the hypothesis since we do not take any steps on the left program.

To overcome these two limitations, we utilize presampling tapes which we saw previously and \emph{error credits}, a separation logic resource first introduced in Eris~\cite{eris}. Error credits are written with the $\upto{\err}$ assertion, where $\err$ is a non-negative real. Intuitively, ownership of $\upto{\err}$ error credits means we are attempting to prove a goal allowing some error of at most $\err$ probability. In \theaplog, error credits are mainly used for a proof technique called \emph{induction by error amplification}, specifically in examples where it is easier to prove that the two programs are equivalent by showing that the distance between them is bounded by errors $\err$ for all $\err>0$, like this rejection sampler.

We now review the basic rules of error credits.
The rule \ruleref{err-split} states that we can split error credits by addition. We can derive $\FALSE$ with $\upto{1}$ error credits, which holds because any proposition, including $\FALSE$, holds with probability at least $0$ (\ruleref{err-1}). We can generate some arbitrarily-small positive error credits from thin air with \ruleref{pupd-err}. Finally, the error amplification rule \ruleref{ind-err-amp} allows us to prove any proposition if we are able to amplify some positive error credit by an amplification factor $k>1$.
\smalldisplay{\begin{mathpar}
  {
    \inferHB{err-split}{\upto{\err_1 + \err_2}}{\upto{\err_1} \sep \upto{\err_2}}
  }
  \and
  \inferH{err-1}
  {\upto{1}}
  {\FALSE}
  \and
  \infrule[lab]{pupd-err}
  {}
  { \pupd{(\Exists \err. 0<\err \sep \upto{\err})} }
  \and
  \inferH{ind-err-amp}
         {
           \begin{matrix}
             \err>0 \qquad k>1 \qquad \upto{\err}\qquad \cr
             (\upto{k\cdot\err} \wand \prop)\wand \upto{\err}\wand \prop
             \end{matrix}
         }
         {\prop}
\end{mathpar}}%
The right-to-left direction of the refinement is established via an intermediate program, similar to the right-to-left refinement of the batch sampling example. In particular our intermediate program is one that matches the ideal $\Rand\tapeboundB$ program but is modified to use tapes, i.e.~$\Fun  \_ . \Rand \ghostcode{(\AllocTape\tapeboundB)}~\tapeboundB$. It is easy to show that the original $\Rand\tapeboundB$ program refines this intermediate program with \ruleref{ht-couple-rand-lbl}. Now, for the refinement between the intermediate program and the rejection sampler program, we first apply \ruleref{pupd-err} to generate some arbitrarily-small error credit. Then we apply \ruleref{ind-err-amp}, choosing $k\eqdef\frac{\tapebound+1}{\tapebound-\tapeboundB}>1$. Subsequently we apply \ruleref{pupd-couple-fragmented}, a rule similar to \ruleref{ht-couple-fragmented} except that it couples a tape presampling operation on the left and a $\Rand\tapebound$ on the right, and it amplifies some error credit by the amplification factor $k$ in the case that we sample a rejected value on the right (we take $f$  to be the identity function).
\smalldisplay{\[
  \inferH{pupd-couple-fragmented}
         { \tapeboundB<\tapebound\\
           \tid\tpmapsto \fillctx \lctx [\Rand\tapebound]
           \\
           \progtape{\lbl}{\tapeboundB}{\tape}\\
           \upto{\err}\\
           f~\text{injective}}
         { \pupd \Exists n\leq\tapebound.
           \tid\tpmapsto \fillctx \lctx [n]\sep 
           \left( \begin{array}{l}
             (\Exists m\leq\tapeboundB. f(m)=n\sep\progtape{\lbl}{\tapeboundB}{\tape\lapp[n]}
           )
           \lor\cr
           \neg(\Exists m\leq\tapeboundB. f(m)=n) \sep \progtape{\lbl}{\tapeboundB}{\tape}
           \sep
           \upto{\frac{\tapebound+1}{\tapebound-\tapeboundB}\cdot\err}
           \end{array}\right)
         }
         \]}%
The proof then proceeds in a straightforward way after a case split on the result of $\Rand\tapebound$. If it is accepted,  we directly read out the sampled value in the tape on the left-hand side with \ruleref{ht-rand-tape}. Otherwise, we step the right-hand side program back to the start of the recursive function, and apply the induction hypothesis of  \ruleref{ind-err-amp}, where we  returned to where we started while successfully amplifying the error by $k$.

To summarize, for programs exhibiting a rejection sampling pattern, we use fragmented couplings to couple the rejection sampling with the simplified ideal one and apply induction by error amplification to show approximate equivalence of the programs for any positive error $\err$. This brings us to our third slogan below that captures the key ideas of proving rejection samplers in \theaplog.
\begin{minipage}{\textwidth}
  \begin{result}
    \begin{center}

\textbf{\hypertarget{slogan3}{Slogan 3}: Fragmented couplings and Error Amplification for Rejections, a.k.a. FEAR.  }
    \end{center}
\end{result}
\end{minipage}

\section{Case Studies}
\label{sec:case-studies}
In this section, we present more complex case studies that demonstrate the use of the advanced features of  \theaplog featured in \cref{sec:logic} and highlight many subtleties in reasoning about concurrent probabilistic programs. Results in this section are out of scope of previous techniques since they involve a combination of concurrency, probability, and higher-order functions.

\subsection{Adversarial von Neumann Coin}
\label{sec:vncoin}
John von Neumann~\cite{neumann-coin} showed that a biased coin can be used to simulate the result of  flipping a fair coin. This is done by repeatedly tossing the coin twice until the results of a pair of tosses are different, at which point we return the result of the first toss. 
\citet{pcol} showed that this result is also true if the bias is stored in shared memory and can be modified as long as it satisfies some invariant, and only fetched right before a pair of tosses. Informally, this means that the program still behaves like a fair coin even when composed with \emph{any} other terminating threads. We formally show that this claim holds by showing contextual equivalence between the following von Neumann coin program and a simple program that samples from $\Rand 1$ (see \cref{fig:vonneumann}).

\begin{figure}[ht!]
  \small
  \begin{minipage}[c]{0.4\linewidth}
    \begin{align*}
      \Lam \adv. & \Let l = \Alloc 0 in \Fork (\adv~l);\\
      &\left(
      {\begin{array}{l}
          \Rec f~\_ = \Let b= \minprog~(\deref l)~\tapebound in\\
          \spac \Let x = \Rand (\tapebound+1) \leq b in \\
          \spac \Let y = \Rand (\tapebound+1) \leq b in \\
          \spac \If x = y then f~\TT \Else x
    \end{array}}
      \right)
    \end{align*}
    \end{minipage}
  \begin{minipage}[c]{0.1\linewidth}
    \begin{align*}
      &\ctxeqrel
      \end{align*}
    \end{minipage}
  \begin{minipage}[c]{0.1\linewidth}
    \begin{align*}
      &\Fun  \adv~\_ . \Flip
    \end{align*}
  \end{minipage}
  \caption{Adversarial von Neumann coin example.}
  \label{fig:vonneumann}
\end{figure}

Both programs in \cref{fig:vonneumann} first take in an adversary argument $\adv$ of type $\tref\tnat\ra\tunit$ and return a closure. Specifically for the adversarial coin program, it creates a reference $l$ containing $0$ and forks a thread that calls $\adv$ with it. This process allows $\adv$ to modify the reference to any natural number in any way, which might involve forking more threads or modifying the reference according to some probabilistic result. The closure returned by the adversarial coin program is a recursive function that sets $b$ to be the minimum of $\tapebound$ and the value read from $l$. This represents setting the bias of the coin to be $(b+1)/(\tapebound+1)$, and we flip the coin twice, and return the first result if they differ. If they are the same, we call the closure again.

Most of this proof follows similarly to that of the rejection sampler example seen before (\hyperlink{slogan3}{\textbf{Slogan 3}} in \cref{sec:advanced-rules-foxtrot}) albeit a more complicated fragmented coupling rule. In this example, we also have to reason about unknown adversarial code $\adv$, which we do so with the \emph{unary} logical relations introduced in \cref{sec:logical-relations}. When proving the left-to-right refinement,  we apply \cref{thm:fundamental-unary} to symbolically execute $\adv~l$ in the forked thread on the left. Note that we \emph{cannot} apply the regular binary logical relations to step $\adv~l$ since we cannot couple the adversarial code with anything on the right. The right-to-left direction is even simpler; after forking $\adv~l$, the angelic nondeterminism of the right-hand side program enables us to directly discard the newly-forked thread resource and focus  only on the expression in the main thread.

\subsection{Sodium}
\newcommand{\MAX}{\textlog{MAX}}
\newcommand{\minv}{\textlog{min}}
Sodium~\cite{libsodium} is a secure cross-platform cross-language open-source cryptography software library. We show the correctness of the \textlog{randombytes\_uniform} function in the library for generating uniform samples (see \cref{fig:sodium}). Our result is particularly strong because it proves that the function behaves as intended even in contexts that might involve \emph{concurrency}, thus showing that the function is \emph{thread-safe} even though its implementation uses a rejection sampling technique and that the random sampling is not atomic (though it appears logically atomic to an outside observer). 

The Sodium function first takes in an argument $\tapebound$. We modify our programs slightly by adding in an additional check that $\tapebound$  is smaller than $\MAX\eqdef{}2^{32}$, which is not needed in the original C implementation since  the argument is an unsigned integer of $32$ bits. We then perform a simple check on whether $\tapebound$ is smaller than $2$ and we simply return $0$ if that is the case. Otherwise, we take $\min$ to be $\MAX~\textlog{mod}~\tapebound$ and we repeatedly sample a random number of 32 bits until it is larger or equal to $\min$\footnote{In the actual implementation, this sampling of a random number of $32$ bits is done by reading from the \textbf{/dev/random} file, which we model directly with a $\Rand$.}, and we eventually return the remainder of the number divided by $\tapebound$. We show that this somewhat convoluted implementation of a rejection sampler is contextually equivalent to a function that samples from $\{0,\dots,\tapebound-1\}$ directly.

\begin{figure}[ht!]
  \small
  \begin{minipage}[c]{0.4\linewidth}
    \begin{align*}
      \Lam \tapebound. & \If \MAX \leq \tapebound then 0 \\
      &\Else \If \tapebound < 2 then 0\\
      &\Else \Let \minv = \MAX~\textlog{mod}~\tapebound in \\
      & \spac \Let r = \Alloc~0 in \\
      & \spac \left(
      {\begin{array}{l}
          \Rec f~\_ = r\gets \Rand (\MAX - 1);\\
          \spac \If \deref r < \minv then f~\TT  \Else (\deref r~\textlog{mod}~\tapebound)
    \end{array}}
      \right)~\TT
    \end{align*}
    \end{minipage}
  \begin{minipage}[c]{0.1\linewidth}
    \begin{align*}
      &\ctxeqrel
      \end{align*}
    \end{minipage}
  \begin{minipage}[c]{0.1\linewidth}
    \begin{align*}
      \Lam \tapebound. &\If (\MAX\leq\tapebound||\tapebound = 0)\\
      & then 0 \\
      & \Else \Rand (\tapebound -1)
    \end{align*}
    \end{minipage}  
  \caption{Sodium \textlog{randombytes\_uniform} implementation.}
  \label{fig:sodium}
\end{figure}

The main bulk of the proof can be split into two stages and it follows very similarly to the arguments we have seen in the batch-sampling and  rejection sampler example in \cref{sec:advanced-rules-foxtrot}. In the first stage, we show that the implementation of the Sodium function is contextually equivalent to one that samples an integer $n$ from $\Rand~(\MAX-(\MAX~\textlog{mod}~\tapebound)-1)$ and returns $n~\textlog{mod}~\tapebound$. This is done by applying the fragmented coupling rules captured in \hyperlink{slogan3}{\textbf{Slogan 3}}  for reasoning about rejection sampling. Then, in the second stage, we show that this intermediate program $p_1$ is contextually equivalent to the idealized function that samples from $\Rand(\tapebound-1)$ directly. This equivalence is established via another intermediate program $p_2$ that concurrently samples from $\Rand(\tapebound-1)$ and $\Rand (\MAX/\tapebound-1)$ with two threads, but only returns the result of the first thread. For the equivalence between $p_1$ and $p_2$, we apply the advanced coupling rules captured in \hyperlink{slogan2}{\textbf{Slogan 2}} to couple the results of two threads with one.  Finally, we show that $p_2$ is contextually equivalent to the idealized program directly with the regular coupling rule \ruleref{ht-couple}.

\subsection{Other Case Studies}
\label{sec:other-case-studies}
For reasons of space, other case studies demonstrating the flexibility of our approach can be found in 
\ifbool{fullversion}{
  the Appendix
}{
  the extended version of this paper~\cite{foxtrot-extended}
}. For example, similar to a result in \citet{prob-nondeterminismLR}, we show that there exist programs which are contextually equivalent to one another even though there are no optimal schedulers that witness the equivalence\ifbool{fullversion}{ (\appref{app:optimal-scheduling})}{}.
We also present a concurrent one-time pad example that is beyond the scope of all previous techniques,  where presampling tapes are utilized  to linearize concurrent random samplings\ifbool{fullversion}{ (\appref{app:one-time-pad})}{}. 
Finally, we study the equations of the algebraic theory induced by the contextual equivalence relation in \theaplog\ifbool{fullversion}{ (\appref{app:algebraic-theory})}{}.

\section{Semantic Model and Soundness}
\label{sec:model}
In this section, we describe the semantic model of \theaplog, defined in the Iris base logic~\cite{irisjournal}, and the key ideas of its soundness proof.

\subsection{Full-information Schedulers}
\label{sec:fischedulers}
As described in \cref{sec:operational-semantics}, since \thelang is concurrent, we resolve the non-determinism of scheduling the threads through a (stateful probabilistic) scheduler. However, the definition of a scheduler in \cref{sec:operational-semantics} is not easy to work with directly when constructing the model of \theaplog. Here, we introduce an alternative  scheduler called a \emph{full-information scheduler} that is only used internally for defining the logic, which we simply write as FIsch(es). We write $\FIsch$ to denote the type of FIsches.

\paragraph{Basic Definition of FIsches}
From a high-level perspective, FIsches are different from normal schedulers in three aspects.
Firstly, the internal state of FIsches are fixed to be $\fullinfostate\eqdef{}List (\Cfg'\times\nat)$, where $\Cfg'\eqdef{}\List(\Expr) \times \Heap$. We implicitly coerce configurations $\cfg$ of type $\Cfg$ to that of type $\Cfg'$ by removing the tape map ($\Tapes$)\footnote{We remove the tape map to ensure FIsches cannot schedule threads based on the content of the tapes.}. Given two FIsch states $\schstate, \schstate'\in\fullinfostate$, we write $\schstate\leq\schstate'$ if $\schstate$ is a prefix of $\schstate'$.

Secondly, the type of FIsches is different from a normal scheduler. Specifically, FIsches are defined by a transition function $\fisch$ of the type $(\fullinfostate \times \Cfg) \to \toption(\Distr{\nat})$.

Lastly, all FIsches $\fisch$ satisfy a \emph{consistency} condition, where for all FIsch states $\schstate$ and configurations $\cfg$, if  $\fisch(\schstate,\cfg)=\None$, then for all other configurations $\cfg'$ we have $\fisch(\schstate,\cfg')=\None$.

\paragraph{Semantics of FIsches}We modify the operational semantics of \thelang with respect to FIsches, such that the thread-stepping function $\fitpstepdistr$ of FIsches is more ``eager'', i.e.~ it attempts to step the corresponding thread regardless of whether the  configuration is final or not:
\smalldisplay{\[
  \fitpstepdistr(\vec \expr, \state)(\threadid) \eqdef
  \begin{cases}
    \mret (\vec \expr, \state) & \text{if \(\expr_\threadid \in \Val\) or $j \geq |\vec \expr|$ ,}\\
    \stepdistr(\expr_\threadid, \state) \mbindi
    \Lam(\expr_\threadid', \state', \vec \expr').
    \mret (\vec \expr [\threadid \mapsto \expr_\threadid'] \dplus \vec \expr' , \state')
      & \text{otherwise.}
  \end{cases}
\]}%
Given a configuration $\cfg$, a FIsch $\fisch$, and a FIsch state $\schstate$, we can now define the single FIsch-step reduction function $\fischstepdistr{\fisch}{\schstate,\cfg} \in \Distr{\Schstate\times\Cfg}$ as follows:
\smalldisplay{\begin{align*}
  \fischstepdistr{\fisch}{\schstate,\cfg} \;\eqdef
  \begin{cases}
    \distr \mbindi \Lam j. \fitpstepdistr(\cfg, j) \mbindi \Lam \cfg'. \mret (\schstate\lapp[(\cfg,j)], \cfg') & \fisch (\schstate,\cfg) = \Some \distr
    \\
    \mret (\schstate, \cfg) & \text{otherwise}
  \end{cases}
\end{align*}}%
Here, in addition to stepping the configuration  according to the FIsch $\fisch$, $\fischstepdistrp$ also updates the FIsch's internal state by extending it with the configuration $\cfg$ and the thread $j$ it attempts to step. Intuitively, this captures the idea that FIsches are keeping track of all the information it has access to since the beginning of the execution, hence the term ``full-information''.

The $n$-step program execution of a FIsch $\fisch$ returns both its internal state and the configuration if the FIsch reaches a $\None$ state:
\smalldisplay{ \begin{align*}
  \fiexec_{\fisch, n}(\schstate, \cfg) &=
                                      \begin{cases}
                                        \mret(\schstate,\cfg) & \fisch(\schstate,\cfg) = \None\\
                                        \nulldistr & n=0\\
                                        \fischstepdistr{\fisch}{ \schstate, \cfg} \mbindi \fiexec_{\fisch, n-1} & \text{otherwise.}
                                      \end{cases}
\end{align*}}%
In contrast to the $n$-step program execution of a normal scheduler, $\fiexec_{\fisch, n}$ returns a distribution on the FIsch state and configuration, instead of a distribution of values. Intuitively, $\fiexec_{\fisch, n}$ simulates the $n$ steps of the execution of a FIsch $\fisch$ until it returns a $\None$.
The full program execution of a FIsch is taken as the limit of its $n$-step program execution, i.e.~$\limfiexec_{\fisch}(\schstate, \cfg) \eqdef \lim_{n\rightarrow \infty} \fiexec_{\fisch, n} (\schstate, \cfg)$.

\paragraph{Properties of FIsches}
As mentioned, FIsches are only used internally to define the model of \theaplog. We relate FIsches and our original definition of schedulers via the following lemma:
\begin{lemma}\label[lemma]{thm:fisch-sch}
  Given a  FIsch $\fisch$, there exists a (regular) scheduler $\sch$ such that for all FIsch states $\schstate$ and values $\val$, we have $(\limfiexec_{\fisch}(\schstate,\cfg)\mbindi g) (\val) \leq\limexecVal_{\sch}(\schstate,\cfg)(\val)$ where \[g\eqdef \Lam (\schstate', \cfg').
  \begin{cases}
    \mret \val& \val\in\Val \land \cfg'.1=\val\lapp\vec\expr \\
    \nulldistr & \text{otherwise}
  \end{cases}\]
\end{lemma}

We now consider various  ways to construct FIsches, which are used in proving the soundness of the model of \theaplog and its proof rules. 

Firstly, the $\liftfisch$ function takes a FIsch $\fisch$ and FIsch state $\schstate$ as input, and returns a  FIsch that behaves the same as $\fisch$ but as if all states are additionally prepended with $\schstate$  in front.
\begin{lemma}\label[lemma]{lem:lift-fisch}
  There exists a function $\liftfisch$ such that given FIsch $\fisch$ and FIsch state $\schstate$,
   $\liftfisch(\schstate,\fisch)$ is a FIsch where for all scheduler state $\schstate'$ and configurations $\cfg$, we have $\limfiexec_{\liftfisch(\schstate,\fisch)} (\schstate\lapp\schstate', \cfg)= \fisch(\schstate', \cfg)\mbindi \Lam (\schstate'',\cfg'). (\schstate\lapp\schstate'',\cfg')$.
\end{lemma}

Using $\liftfisch$, we can define the $\appfisch$ function, allowing us to chain a FIsch with a family of FIsches. Intuitively, $\appfisch(\fisch, f)$ is a FIsch that first acts like $\fisch$ until we reach a state $\schstate_2$ for which $\fisch$ always returns $\None$, at which point, we continue to behave as the Fisch $f(\schstate_2)$. \diffred{Later in the proof of the adequacy theorem (\cref{thm:adequacy}), this $\appfisch$ function is our main weapon in piecing various FIsches together (see \cref{sec:soundness}). }
\begin{lemma}\label[lemma]{lem:app-fisch}
  There exists a function $\appfisch$ such that
  given FIsch $\fisch$ and a function $f$ of type $\fullinfostate\ra\FIsch$,  $\appfisch(\fisch, f)$ is a  FIsch where for all $\cfg, x$, we have
  $(\limfiexec_{\fisch}(\nil,\cfg)\mbindi g) (x) \leq \limfiexec_{\appfisch(\fisch, f)} (\nil,\cfg)(x)$
  where \[g(\schstate_1,\cfg_1) \eqdef 
  \begin{cases}
	  \limfiexec_{\liftfisch(\schstate_2, f(\schstate_2))} (\schstate_1, \cfg_1) & \text{exists}~\schstate_2\leq\schstate_1~\text{minimal s.t.}~\forall \rho'. \fisch(\schstate_2,\rho')=\None \\
    \nulldistr & \text{otherwise}
  \end{cases}
  \]
\end{lemma}

\subsection{Semantic Model}
\label{sec:semantic-model}
\paragraph{Weakest Precondition}
We follow the standard trick of expressing Hoare triples in terms of a weakest precondition~\cite{irisjournal}, i.e. $\hoare{\prop}{\expr}{\propB}\eqdef{}\prop\wand\wpre{\expr}{\propB}$. 
 The definition of the weakest precondition (shown below) follows similar structure as previous logics~\cite{approxis, coneris} where it is defined as a guarded fixed point (all recursive occurrences of the weakest precondition appear under the later modality $\later$). We emphasize that the novelty of the \theaplog weakest precondition lies in the significant differences in how the spec-coupling ($\specStepp$) and program-coupling ($\progStepp$) preconditions are defined, which we detail later. 
 \smalldisplay{
   \begin{align*}
  \wpre{\expr_{1}}{\pred} \eqdef{}
  & \All \state_{1}, \cfg_{1}, \err_{1} .
    \stateinterp~\state_{1}~\cfg_{1}~\err_{1} \wand \pvs[\top][\emptyset]\specStepl{\state_{1}}{\cfg_{1}}{\err_{1}}\spac\{ \state_{2}, \cfg_{2}, \err_{2} \ldotp \\
  & \quad \big(\expr_{1} \in \Val \sep\pvs[\emptyset][\top]
    \stateinterp~\state_{2}~\cfg_{2}~\err_{2} \sep \pred~\expr_{1}\big) \lor{} \\
  & \quad
    \big( \expr_{1} \not\in \Val \sep
    \progStepl{(\expr_{1}, \state_{2})}{\cfg_{2}}{\err_{2}}
    \spac\{ \expr_{2}, \state_{3}, l, \cfg_{3}, \err_{3} \ldotp  \\
  & \qquad \later \specStepl{\state_{3}}{\cfg_3}{\err_{3}}
    \spac\{ \state_{4}, \cfg_{4}, \err_{4} \ldotp 
    \pvs[\emptyset][\top]\stateinterp~\state_{4}~\cfg_4~\err_{4} \sep \wpre{\expr_{2}}{\pred} \sep
    \textstyle\Sep_{\expr'\in l} \wpre{\expr'}{\TRUE}
    \} \} \big) \}
\end{align*}}%
To prove a weakest precondition, we initially assume the ownership of the \emph{state interpretation} $\stateinterp~\state_{1}~\cfg_{1}~\err_{1}$, which interprets the physical state of the implementation program, the specification program, and the approximate error as resources in the standard way~\cite{irisjournal, approxis, coneris}. It hence gives meaning to various resources e.g.~the points-to connective $\loc\mapsto\val$, the specification thread connective $\tid\tpmapsto\expr$, and the error credits $\upto{\err}$.
We then get access to the resources in all invariants through an update modality $\pvs[\top][\emptyset]$ and we need to prove a \emph{spec-coupling} precondition $\specStepl{\state_{1}}{\cfg_{1}}{\err_{1}}\{\dots\}$. We explain its definition later in \cref{fig:specStep}; for now it suffices to interpret it as a precondition for allowing the right-hand side  program to progress.

We subsequently perform a case split on whether $\expr_1$ is a value. If it is, we do a view shift $\pvs[\emptyset][\top]$ where we re-establish all invariants, return the updated state interpretation and prove that $\expr_1$ satisfies the postcondition $\pred$. Otherwise, we prove a \emph{program-coupling} precondition $\progStepl{(\expr_{1}, \state_{2})}{\cfg_{2}}{\err_{2}}\{\dots\}$. As before, we explain the details of this precondition later, but for now one can think of it as a precondition similar to that of $\specStepp$, but it allows us to also take an actual step on the configuration $(\expr_1, \state_2)$ to obtain resulting expression $\expr_2$, state $\state_3$, a list of forked expressions $l$, and resulting error budget $\err_3$. We then prove another $\specStepp$ precondition \diffred{(which can be ignored here and is only needed to establish \ruleref{ht-inv-open})}. We finally re-establish all invariants through the view shift $\pvs[\emptyset][\top]$ and return the newly updated state interpretation, prove that $\wpre{\expr_2}{\pred}$ holds, and show that all forked expressions are safe to execute, i.e.~$\wpre{\expr'}{\TRUE}$ for every expression $\expr'$ in the list $l$.

\paragraph{Coupling Preconditions}
We now turn our attention to the spec-coupling ($\specStepp$) and program-coupling ($\progStepp$)  preconditions. \diffred{At a high level, these preconditions are defined with FIsches that satisfy coupling properties, which then become the building blocks to be ``pieced together'' in the proof of the adequacy theorem (\cref{thm:adequacy}).  }

We first present two auxiliary definitions.
Firstly, we say a relation $R\subseteq A \times (B\times C)$ is state injective, written as $\stateinjective(R)$, if it is functional in $B$, i.e.~for all $a, a', b, c, c'$, if $R~a~(b,c)$ and $ R~a'~(b,c')$, then $a=a'$ and $c=c'$.

Secondly, we define the notion of a scheduler erasable state update, which intuitively describes actions on the state (mostly on the tape heap) that do not alter the operational semantics of the expressions; examples of scheduler erasable state updates include tape presamplings actions.
\begin{definition}
  A distribution on states $\distr$ is a \defemph{scheduler erasable state update} of $\state \in \State$, written as $\scherasable (\distr,\state)$, if for all schedulers $\sch$, scheduler states $\schstate$, thread pools $\vec e$, and integers $n$,
  we have
  \begin{align*}(\distr \mbindi (\Lam \state' . \pexec_{\sch, n} (\schstate, (\vec e, \state')))).\mathsf{tp} \;=\; (\pexec_{\sch, n} (\schstate, (\vec e, \state))).\mathsf{tp}
  \end{align*}
  where the function \(-.\mathsf{tp}\) projects out the thread pool component from a distribution on configurations.
\end{definition}
The spec-coupling precondition $\specStepp$ is defined inductively by four inference rules shown in \cref{fig:specStep}. If the error budget $\err$ is lower-bounded by $1$, the precondition holds trivially as all sub-distributions have mass smaller or equal to $1$ (\ruleref{spec-step-err-1}). If the postcondition holds for the input parameters, the precondition also holds trivially (\ruleref{spec-step-ret}). If the precondition holds for any error budget $\err'$ larger than $\err$, then the precondition also holds for $\err$ (\ruleref{spec-step-continuous}). This rule enables us to prove \ruleref{pupd-err} and create error credits from thin air.

We now focus on \ruleref{spec-step-exp}, which is the most important inference rule in the $\specStepp$ precondition. Intuitively, this rule establishes an $\err'$-approximate coupling between a scheduler erasable state update $\mu$, e.g.~a tape presampling action on the left, with some execution steps of the right-hand side specification program under some state injective relation $R$. Here, the execution of the specification program is captured by the limit execution of a FIsch $\fisch$ with the empty list $\nil$ as the starting state. In addition, the rule allows us to distribute the remaining error according to some function $\Err$ as long as it satisfies an expectation-preserving inequality.
\begin{figure*}[]
  \small
  \centering
  \begin{mathpar}
    \inferH{spec-step-err-1}
    {1\leq\err}
    { \specStep{\state}{\cfg}{\err}{\Phi} }
    \and
    \inferH{spec-step-ret}
    { \Phi(\state,\cfg,  \err) }
    { \specStep{\state}{\cfg}{\err}{\Phi} }
    \and
    \inferH{spec-step-continuous}
    { \All \err'.\err<\err' \wand \specStep{\state}{\cfg}{\err'}{\Phi} }
    { \specStep{\state}{\cfg}{\err}{\Phi} }
    \and
    \inferH{spec-step-exp}
    {
      \scherasable(\mu, \state_1) \\
      \Exists r. \All x. \Err_2 (x)\leq r\\
      (\err' + \expect[\limfiexec_{\fisch}(\nil, \cfg_1)]{\Err}) \leq \err \\
      \ARcoupl{\mu}{\limfiexec_{\fisch}(\nil, \cfg_1)}{\err'}{R}\\
      \stateinjective(R)\\
      \All \state_2,  (\schstate, \cfg_2).
      R~\state_2~(\schstate,\cfg_2)  \wand
      \specStep{\state_2}{\cfg_2}{(\Err(\schstate, \cfg_2))}{\Phi}
    }
    { \specStep{\state_1}{\cfg_1}{\err}{\Phi} }
  \end{mathpar}
  \caption{Inductive Definition of the spec-coupling precondition $\specStep{\state}{\cfg}{\err}{\Phi}$.}
  \label{fig:specStep}
\end{figure*}

The program-coupling precondition $\progStepp$ is defined by the single inference rule \ruleref{prog-step-exp} which is of a similar structure as that of \ruleref{spec-step-exp}. The main difference here is that this rule couples a single execution step on the left-hand side program with some execution steps on the right-hand side program according to some FIsch $\fisch$. (The proposition $\red(\expr,\state)$ means that the configuration $(\expr,\state)$ is reducible for one execution step.)
\smalldisplay{\begin{mathpar}
  \inferH{prog-step-exp}
  { \Exists r. \All x. \Err (x)\leq r\\
    (\err' + \expect[\limfiexec_{\fisch}(\nil, \cfg_1)]{\Err}) \leq \err \\
    \ARcoupl{\stepdistr (\expr_1, \state_1)}{\limfiexec_{\fisch}(\nil, \cfg_1)}{\err'}{R}\\
     \red(\expr_1, \state_1) \\
    \stateinjective(R)\\
     \All (\expr_2, \sigma_2, l), (\schstate, \cfg_2) . R~(\expr_2, \sigma_2, l)~(\schstate, \cfg_2)\wand
     \Phi (\expr_2, \state_2, l, \cfg_2, \Err(\schstate, \cfg_2)) }
  {\progStep{(\expr_1, \state_1)}{\cfg_1}{\err}{\Phi}}
\end{mathpar}}%
\paragraph{Probabilistic Update Modality}
We follow the approach of Coneris~\cite{coneris} in defining the probabilistic update modality. Just like the fancy update modality $\pvs[\mask_1][\mask_2]$, the probabilistic update modality is in fact additionally annotated with two sets of invariants, i.e.~$\pupd[\mask_1][\mask_2]$.
Intuitively, $\pupd[\mask_1][\mask_2]\prop$ is defined by first assuming ownership of some state interpretation, opening all invariants in $\mask_1$, proving a spec-coupling precondition with  input parameters $\state_1$, $\cfg_1$, and $\err_1$, re-establishing all invariants in the mask $\mask_2$, and finally giving back the state interpretation and  resource $\prop$.
\smalldisplay{\begin{align*}
  \pupd[\mask_1][\mask_2] P \eqdef{}
  &\All \state_{1}, \cfg_1, \err_{1} . 
  \stateinterp~\state_{1}~\cfg_1~\err_{1}~ \wand \pvs[\mask_1][\emptyset]\specStepl{\state_{1}}{\cfg_1}{\err_{1}}\spac\{ \state_{2}, \cfg_2, \err_{2} \ldotp
  \pvs[\emptyset][\mask_2]\stateinterp~\state_{2}~\cfg_2~\err_{2} \sep P \}
\end{align*}}

\subsection{Soundness}
\label{sec:soundness}
As hinted in \cref{sec:key-ideas}, to compose families of FIsches, we rely on an Iris version of the axiom of choice:
\begin{lemma}[Iris Choice]
  \label[lemma]{thm:choice}
  We have $(\All a. \Exists b. \prop~a~b) \vdash \Exists f. \All a. \prop~a~(f(a))$ within the Iris logic~\footnote{Assuming the axiom of choice in the meta-logic.}.
\end{lemma}
Perhaps this result might appear suspicious at first glance. \citet{diaconescu} showed that the original axiom of choice implies the law of excluded middle, which is known to not hold in the intuitionistic Iris separation logic~\cite{iris}. 
The seeming contradiction is avoided by restricting \cref{thm:choice} to only apply when the variables $a$ and $b$ are quantified over meta-level Rocq types. Consequently, the types of $a$ and $b$ cannot utilize the step-indexing and resources of the Iris logic, as would be necessary to replay Diaconescu's proof.

The  adequacy theorem (\cref{thm:adequacy}) is proven via the following intermediate lemma \cref{thm:adequacy-simpl}\footnote{We quantify over errors $\err'>0$ since sometimes an optimal scheduler cannot be found for the
  right-hand side program, but we can find one that approximately couples with
  the left for any  positive error (see \appref{app:optimal-scheduling}).}. We then recover  \cref{thm:adequacy} by first taking $\vec\expr$ to be the empty list $\nil$, applying \cref{thm:fisch-sch} to eliminate the use of FIsches, and finally taking the limit of $n$. 

\begin{lemma}
  \label[lemma]{thm:adequacy-simpl}
  If we can prove \[\upto{\err} \sep \Sep_{(\tid,\expr')\in \textlog{enum}(\vec {\expr_{s}})}\tid\tpmapsto\expr'\vdash \wpre{\expr}{\val.\Exists \val'.0\tpmapsto\val'\sep\pprop~\val~\val'} \sep \Sep_{\expr'\in \vec e} \wpre{\expr'}{\TRUE}\] in \theaplog,
  then for all schedulers $\sch$, scheduler states $\schstate$, states $\state, \state'$, natural numbers $n$, and error $\err'>0$,
  there exists a  FIsch $\fisch$ such that $\ARcoupl{\execVal_{\sch,n}(\schstate, (\expr\cons\vec \expr,\state))}{\limfiexec_{\fisch}(\nil,(\vec{\expr_{s}},\state'))}{\err+\err'}{\ppropB}$ where
\[\ppropB~\val~(\schstate',\cfg')\eqdef\Exists \val'~\vec \expr'.\cfg'.1=\val'\cons\vec\expr'\land \pprop~\val~\val'\]
  
\end{lemma}
Intuitively, \cref{thm:adequacy-simpl} states that for any scheduler $\sch$ on the left and any  execution steps $n$, we can find a FIsch $\fisch$ for the right such that we can couple the $n$-step execution of the left-hand side program with the limit execution of the one on the right. In the proof of \cref{thm:adequacy-simpl}, we start by fixing the arbitrary scheduler for the left-hand side program. 
\diffred{Recall that the preconditions $\progStepp$ and $\specStepp$ are defined with FIsches for the right-hand side program that satisfy various coupling properties.
For each step of the program on the left, unfolding the weakest precondition of \theaplog (specifically the spec-coupling and program-coupling preconditions) gives us various FIsches for the right and an associated probabilistic coupling property for each. The key idea of the proof is to ``piece together'' all the FIsches and couplings from $\progStepp$ and $\specStepp$ to produce a single FIsch for the right-hand side program, along with a coupling that covers the entire execution.}

\diffred{This piecing operation is done  by induction on $n$; given a FIsch that couples the first step and a family of FIsches that couples $n$ more steps for each possible outcome after the first step, we can apply \cref{thm:choice} and the $\appfisch$ function to glue the first FIsch with the family of FIsches to form a new FIsch that couples the first $n+1$ steps of execution. Here \cref{thm:choice} is needed because the size of the family of FIsches might be countably infinite.
Interestingly, instead of performing a \emph{single} ``FIsch-piecing'' operation  in the inductive step, we actually have to glue the FIsches a total of $4$ times in the inductive step! We provide a more detailed proof sketch of \cref{thm:adequacy-simpl}  in \appref{app:adequacy-proof}. }

\section{Related Work}
\label{sec:related-work}

\paragraph{Contextual Equivalences of Higher-Order Programs}
There are various prior studies on utilizing binary logical relations to prove contextual equivalences of higher-order programs.
For example, there is a long line of work using step-indexed logical relations for establishing refinements of higher-order concurrent languages. In particular, we highlight  CaReSL~\cite{turon-caresl}, one of the first logics for fine-grained concurrency, which was then mechanized in Iris~\cite{DBLP:conf/popl/KrebbersTB17, aminphd}. The mechanized concurrent separation logic ReLoC~\cite{reloc} extended CaReSL to  internalize refinement judgements as first-class logical statements, providing more concise rules for reasoning about invariants and logical atomicity. Similarly, there are also many logical relations specialized for probabilistic languages. \citet{probabilityLR} defined a step-indexed, biorthogonal logical relation for reasoning about contextual equivalences of a higher-order probabilistic language with discrete random variables and state.  This was then extended for more complicated language features such as continuous random sampling~\cite{culpepper, wandLR} and countable nondeterminism~\cite{prob-nondeterminismLR}. The mechanized logic Clutch~\cite{clutch} first introduced presampling tapes for  asynchronous couplings, which we also use to implement advanced coupling rules for reasoning about rejection sampling and multiple concurrent samplings in \theaplog. We also took inspiration from Approxis~\cite{approxis}, which used error credits to establish contextual equivalences of rejection samplers by means of approximation. 

There are other techniques other than logical relations to verify contextual equivalences, e.g. \citet{pcoh} utilized probabilistic coherence spaces to construct a fully abstract denotational model for PCF programs with a discrete random primitive. Much work has also been done in using bisimulation techniques to establish the contextual equivalences of programs written in the probabilistic $\lambda$-calculus, whether it be call-by-value~\cite{cbv-ce}, call-by-name~\cite{cbname-ce}, or call-by-need~\cite{cbneed-ce}. 

Compared to all the work above, our work in proving contextual equivalences of higher-order (stateful) programs is the \emph{first} to be extended to the concurrent and randomized setting, which we believe is a \emph{significant} contribution as reasoning about this mixture is notoriously hard.

\paragraph{Logics for Probability and Concurrency}
There has been previous work on program logics for reasoning about properties (other than contextual refinement) of concurrent probabilistic programs.

\citet{prob-rely-guarantee} first introduced the probabilistic rely-guarantee calculus for reasoning about the quantitative correctness of probabilistic concurrent programs without support for dynamically-allocated local state. \citet{cqsl} developed concurrent quantitative separation logic for reasoning about quantitative properties of concurrent, randomized, heap-manipulating programs. Polaris~\cite{polaris} is similar to our work in that it is also a mechanized relational logic in Iris, but the refinement is constructed between a concurrent probabilistic program and a simpler monadic representation model, which can then further be analyzed to establish properties of the original program, e.g.~bounds on expected values. ExpIris~\cite{lohse2024iris} is another mechanized program logic in Iris; its Hoare triples are annotated with a parameter called the potential, which is used to prove bounds on expected costs of concurrent probabilistic programs. The recent Probabilistic Concurrent Outcome Logic~\cite{pcol}, a variant of Outcome Logic~\cite{outcomelogic}, supports compositional reasoning  of the distribution of possible outcomes from the execution of concurrent randomized programs.

Most similar to our logic is that of Coneris~\cite{coneris}, a separation logic for proving error bounds of concurrent probabilistic programs. Firstly, the underlying language \thelang of \theaplog is identical to that in Coneris. Moreover, our use of error credits is heavily inspired by that of Coneris~\cite{coneris} (and similar logics such as Eris~\cite{eris} and Approxis~\cite{approxis}), where the error credits were originally used to prove error bounds of (concurrent) probabilistic programs. We also adopt Coneris' probabilistic update modality and extend it into our relational setting for symbolically executing right-hand side programs in addition to presampling onto tapes. \diffred{Instead of proving contextual refinement between two programs, Coneris is a unary logic and is used to establish error probability bounds of individual programs. }

\section{Conclusion}
\label{sec:conclusion}
We presented  \theaplog, the first higher-order separation logic for proving contextual refinements of higher-order concurrent probabilistic programs with higher-order local state. In addition, we demonstrated its strengths on  a wide range of examples involving complex  local state, random sampling, and concurrent behavior, that were previously out of scope.

In the future, we would like to extend \theaplog to reason about contextual refinement of programs under schedulers with restricted power for proving security guarantees of distributed cryptographic protocols, e.g.~schedulers which can only rely on their internal state to decide which thread to step next. We also aim to implement prophecy variables~\cite{prophecy1, prophecy2, iris-prophecy} within \theaplog to prove more interesting equivalences. Lastly, it would be interesting to consider modifying \theaplog for proving must-termination contextual refinements~\cite{nondeterminismLR,prob-nondeterminismLR} of concurrent probabilistic programs.

\section*{Data Availability Statement}
The Rocq formalization accompanying this work is available online on Zenodo~\cite{foxtrot-artifact} and on Github at \href{https://github.com/logsem/clutch}{github.com/logsem/clutch}.

\begin{acks}
  The first author would like to thank Amin Timany and Daniel Gratzer for insightful discussions regarding the axiom of choice.

  This work was supported in part by the \grantsponsor{NSF}{National Science Foundation}{}, grant no.~\grantnum{NSF}{2338317}, a \grantsponsor{Villum}{Villum}{} Investigator grant, no. \grantnum{Villum}{VIL25804} and no. \grantnum{Villum}{VIL73403}, Center for Basic Research in Program Verification (CPV), from the VILLUM Foundation, and the European Union (\grantsponsor{ERC}{ERC}{}, CHORDS, \grantnum{ERC}{101096090}).
  Views and opinions expressed are however those of the author(s) only and do not necessarily reflect those of the European Union or the European Research Council.
  Neither the European Union nor the granting authority can be held responsible for them.
\end{acks}

\bibliography{refs}

\ifbool{fullversion}{
  \pagebreak
  \appendix

\section{Auxiliary Programs}
\label{sec:auxiliary-programs}
We   introduce two auxiliary programs (\cref{fig:nondet-diverge}): $\nondet$ and $\diverge$.

\begin{figure}[ht!]
  \small
  \begin{minipage}[t]{0.45\linewidth}
    \begin{align*}
      \nondet\eqdef{}& \Fun \_ . \\
      &\spac \Let x = \Alloc~0 in\\
      & \spac \Fork ((\Rec f~\_ = x\gets \deref x+1; f~\TT) \TT)\\
      & \deref x
    \end{align*}
    \end{minipage}
  \begin{minipage}[t]{0.45\linewidth}
    \begin{align*}
      \diverge\eqdef{}& \Rec f~\_ = f~\TT
    \end{align*}
    \end{minipage}
  
  \caption{The $\nondet$ and $\diverge$ programs.}
  \label{fig:nondet-diverge}
\end{figure}

The program $\nondet$ is a function that first creates a reference containing $0$. It then spawns a thread that repeatedly increments the reference, and the main thread reads from the reference and returns the value. It is trivial to see that for any natural number $n$, there exists some scheduling such that $\nondet~\TT$ returns $n$, as captured by the rule \ruleref{ht-nondet}. In fact, if this program is on the right-hand side of the refinement, we can deliberately choose the threads to step in a way to return a particular natural number, as captured by \ruleref{pupd-nondet}.

The program $\diverge$ is simply a recursive program that calls itself. With Löb induction, one can prove \ruleref{ht-diverge}, since it never terminates with a value. There is no corresponding rule for $\diverge~\TT$ appearing on the right-hand side of the refinement. 
\smalldisplay{\begin{mathpar}
  \inferH{ht-nondet}
         {}{\hoare{\TRUE}{\nondet~\TT}{v.\Exists (n:\nat). v = n}}
         \and
         \inferH{pupd-nondet}
                {\tid\tpmapsto\fillctx\lctx[\nondet~\TT]}{\pupd{\tid\tpmapsto\fillctx\lctx[n]} \quad (\text{for any natural number $n$})}
                \and
                \inferH{ht-diverge}{}{\hoare{\TRUE}{\diverge~\TT}{\FALSE}}
\end{mathpar}}

\section{Counter example of presampling on the right}
\label{app:counter}
We prove that presampling on the right hand side is unsound  in \theaplog.
Assume we have the following unsound coupling rule, that allows you to couple a $\Rand$ on the left and a presampling on a tape on the right that is sound in previous logics supporting tapes~\cite{clutch, approxis}:
\smalldisplay{\begin{mathpar}
  \inferH{unsound-ht-couple-rand-lbl}
         { 
           \All n \leq \tapebound . \hoare{\spectape{\lbl}{\tapebound}{\tape\lapp[n]}}{n}{\Phi} }
         { \hoare{
           \spectape{\lbl}{\tapebound}{\tape}}
           {\Rand \tapebound}{\Phi} }
\end{mathpar}}
We now define the following four program:
\newcommand{\proga}{\textlog{progA}}
\newcommand{\progb}{\textlog{progB}}
\newcommand{\progc}{\textlog{progC}}
\newcommand{\progd}{\textlog{progD}}
  \smalldisplay{  \begin{align*}
      \proga\eqdef{}& \TT \\
      \progb\eqdef{}& \If \nondet~\TT=\Rand 1 then \TT \Else \diverge~\TT \\
      \progc\eqdef{}& \Let \lbl = \AllocTape 1 in \\
      &\If \Rand \lbl~1=\nondet~\TT then \TT \Else \diverge~\TT \\
      \progd\eqdef{}& \If \Rand 1=1 then \TT \Else \diverge~\TT 
    \end{align*}}%
    Note that programs in \thelang follow a right-to-left evaluation order, so the expression $\nondet~\TT=\Rand 1$ first performs the random sampling before executing $\nondet~\TT$.
    We can prove the following three inequalities:
    \begin{enumerate}
      \item  The upper termination probability  of $\proga$ is upper bounded by that of $\progb$. This is done by  forcing $\nondet~\TT$ to return the same value produced by $\Rand 1$.
      \item  The upper termination probability  of $\progb$ is upper bounded by that of $\progc$. We can prove this with our unsound rule where we perform the coupling between the labelled $\Rand$ on the left and the presampling action on the tape on the specification side. We also force $\nondet~\TT$ to return the value presampled onto the tape. This rest of the proof follows by stepping the programs symbolically.
      \item   The upper termination probability  of $\progc$ is upper bounded by that of $\progd$. This proof is done in a straightforward manner by choosing the right coupling between the two $\Rand$ operations after executing $\nondet~\TT$ on the left. 
    \end{enumerate}

By transitivity, it follows that the upper termination probability  of $\proga$ is upper bounded by that of $\progd$. This is a contradiction since $\proga$ almost surely terminates but $\progd$ does not. 

\section{Example without Optimal Scheduling}
\label{app:optimal-scheduling}
\citet{prob-nondeterminismLR} showed that there are pairs of non-deterministic probabilistic programs that are contextually equivalent with each other even though there does not exist an optimal scheduler that witnesses the equivalence. We present an example showing that this result also extends to the concurrent setting.

Consider the  two contextually equivalent programs in \cref{fig:optimal}. The left-hand side program  first non-deterministically chooses a natural number $n$ and samples from $\Rand n$. If that result is $0$, it loops forever with $\diverge~\TT$, otherwise it directly returns $\TT$. The right-hand side program directly returns $\TT$.
\begin{figure}[ht!]
  \small
  \begin{minipage}[c]{0.33\linewidth}
    \begin{align*}
      &\Let n = \nondet~\TT in \\
      &\If \Rand n=0 then \diverge~\TT\\
      & \spac \Else \TT
    \end{align*}
    \end{minipage}
  \begin{minipage}[c]{0.1\linewidth}
    \begin{align*}
      &\ctxeqrel
      \end{align*}
    \end{minipage}
  \begin{minipage}[c]{0.2\linewidth}
    \begin{align*}
      &\TT
    \end{align*}
    \end{minipage}
  
  \caption{Example without optimal scheduling. }
  \label{fig:optimal}
\end{figure}

What is the least upper bound of the termination probability of both programs? It is obviously $1$ for the right-hand side program, and it turns out it is also $1$ for the left-hand side program as well, even though it never terminates with probability $1$ for \emph{any} scheduling! This is because for any natural number $n$, we can find some scheduling such that the left-hand side program terminates with   probability $1-1/(n+1)$ (by ensuring $\nondet~\TT$ returns $n$). Hence both programs are contextually equivalent! 

The left-to-right direction of the refinement is trivial: after assigning a nondeterministic number to $n$ with \ruleref{ht-nondet}, we proceed via a case distinction on whether the sampled value is $0$. If yes, we can just apply \ruleref{ht-diverge} to obtain a $\FALSE$ hypothesis. Otherwise, both  programs return $\TT$, which satisfy $\semInterpS{\tunit}{\Delta}$.

The right-to-left direction is a bit more involved. Note that we cannot just blindly step through the program because there is no way to continue after arriving at $\diverge~\TT$ on the right-hand side. The trick is to avoid arriving at that branch in the first place, but how can that be done? The key idea is that to use \ruleref{pupd-err} to produce some positive error $\upto{\err}$ at the beginning. Since $\err>0$ there must exists some natural number $n$ such that $\err>1/(n+1)$, and hence we have $\upto{1/(n+1)}$. Now we specifically choose to return $n$ with \ruleref{pupd-nondet}. When we arrive at the $\Rand$ sampling instruction, we use \ruleref{pupd-rand-exp} for distributing error credits across all the possible branches of a $\Rand$ as a weighted sum according to the probability of each branch\footnote{There is similarly a rule for distributing error credits across a $\Rand$ for the left-hand side program as well, which we omit for brevity. }.
\smalldisplay{\[
  \inferH{pupd-rand-exp}
         {\expect[\unifd{N}]{\Err} \leq \err \\
           \upto{\err} \\
           \tid\tpmapsto\fillctx\lctx[\Rand\tapebound]
         }
  { \pupd{\Exists n\leq\tapebound. \tid\tpmapsto\fillctx\lctx[n]\sep \upto{\Err(n)}} }
  \]}%
  Specifically we choose $\Err(n)\eqdef{}[n=0]$. That way, when we do a case distinction on whether the sampled value is $0$, we can get a contradiction with $\upto{1}$ in the case we sampled a $0$. Otherwise, we return $\TT$ for both programs, satisfying $\semInterpS{\tunit}{\Delta}$.

\section{Concurrent Implementation of One-Time Pad}
\label{app:one-time-pad}
  In \cref{sec:advanced-rules-foxtrot}, we used presampling tapes to combine randomness arising from concurrent threads. Here, we verify the correctness of a concurrent implementation of a one-time pad, where we use presampling tapes to not combine the randomness of multiple $\Rand$ operations from concurrent threads, but to linearize them.

  \newcommand{\otpimpl}{\textlog{otpImpl}}
  \newcommand{\otpspec}{\textlog{otpSpec}}
  \newcommand{\otpimplt}{\textlog{otpImpl'}}
  A one-time pad is a simple but powerful encryption technique as it is ``information-theoretically secure'' and cannot be cracked. To encrypt a plaintext, we generate a uniformly-chosen random key and compute the modular addition of the ciphertext and the key. Decryption follows directly by computing the modular addition of the ciphertext and the key.
  
  We consider a simplified program $\otpimpl$ in \cref{fig:otp}, where both the plaintext and key is randomly generated from $\Rand \tapebound$, but done so in a concurrent way, where the two threads perform a fetch-and-atomic-add instruction into a common reference\footnote{Note that the one-time pad is still secure even when the plaintext is not uniformly distributed; we proved a stronger result where the plaintext is generated according to some unknown distribution, and we refer readers to the Rocq repository for more details. }. We prove its security by showing that the program is contextually equivalent to a $\Rand \tapebound$ ($\otpspec$):
  \begin{figure}[ht!]
    \small
  \begin{minipage}[c]{0.4\linewidth}
    \begin{align*}
      &\otpimpl\eqdef{}\\
      &\Let x = \Alloc 0 in \\
  &\left({
    \begin{array}{l}
      \Let \textlog{msg} = \Rand \tapebound in \\
      \Faa x~\textlog{msg}
    \end{array}
    }
    \middle|\middle|\middle|
    {\begin{array}{l}
      \Let \textlog{key} = \Rand \tapebound in \\
      \Faa x~\textlog{key}
    \end{array}}\right); \\
      & \deref x~\text{mod}~(\tapebound+1)
    \end{align*}
    \end{minipage}
  \begin{minipage}[c]{0.1\linewidth}
    \begin{align*}
      &\ctxeqrel
      \end{align*}
    \end{minipage}
  \begin{minipage}[c]{0.1\linewidth}
    \begin{align*}
      & \otpspec\eqdef{}\\
      &\Rand \tapebound
    \end{align*}
    \end{minipage}
  
  \caption{One-time pad example.}
  \label{fig:otp}
\end{figure}

  Unsurprisingly, the right-to-left direction is not too difficult. Since $\otpimpl$ resides on the right-hand side of the refinement, we can take advantage of the angelic nondeterminism of the specification program and choose the interleaving that first resolves the first thread of $\otpimpl$ followed by the second. For the  sampling of the second thread, we couple it with the one in $\otpspec$ via \ruleref{ht-couple}, choosing a bijective function  such that the modular addition results in the same value as that returned by $\otpspec$.

  On the other hand, the left-to-right direction is trickier. Morally, depending on how the threads are scheduled, we want to couple the second $\Rand$ sampling of $\otpimpl$ with the $\Rand$ in $\otpspec$. However, we do not know in advance which thread is going to be scheduled first when the concurrent program $\otpimpl$ is on the left-hand side of our refinement, resulting in a more demonic flavor of nondeterminism.

The key idea for proving the refinement from left-to-right is to linearize the random samplings by using tapes to presample the random samplings in advance before the parallel composition in $\otpimpl$.
Like before, we prove the refinement via  a intermediate program $\otpimplt$, and it is straightforward to show  the refinement between $\otpimpl$ and $\otpimplt$:
\smalldisplay{\begin{align*}
  \otpimplt\eqdef{}&\Let x = \Alloc 0 in \\
  & \ghostcode{\Let \lbl = \AllocTape \tapebound in}\\
  & \ghostcode{\Let \lbl' = \AllocTape \tapebound in}\\
  &\left({
    \begin{array}{l}
      \Let \textlog{msg} = \Rand\ghostcode{\lbl}~\tapebound in \\
      \Faa x~\textlog{msg}
    \end{array}
    }
    \middle|\middle|\middle|
    {\begin{array}{l}
      \Let \textlog{key} = \Rand\ghostcode{\lbl'}~\tapebound in \\
      \Faa x~\textlog{key}
    \end{array}}\right); \\
      & \deref x~\text{mod}~(\tapebound+1)
    \end{align*}}%
It then suffices to prove the refinement between $\otpimplt$ and $\otpspec$, which turns out to be relatively straightforward as well. After allocating the two tape labels in $\otpimplt$, we apply \ruleref{pupd-presample} to $\lbl$, a simple rule for presampling a number onto the tape on the left-hand side program. 
\smalldisplay{\begin{mathpar}
  \inferH{pupd-presample}
  {
    \progtape{\lbl}{\tapebound}{\tape}
  }
  {\pupd
    (\Exists n\leq\tapebound. \progtape{\lbl}{\tapebound}{\tape\cons n})}  
\end{mathpar}}%
Then we apply \ruleref{pupd-couple-lbl-rand} to couple the presampling of the tape $\lbl'$ and the $\Rand$ in $\otpspec$ according to the bijective function such that the modular addition of the two presampled values is the same as the one sampled from the specification program. The proof then proceeds by defining an invariant that shares the reference and captures a protocol of how its stored value changes with respect to the threads (using standard Iris ghost resources). After the threads  sample the previously presampled value from their respective tapes according to \ruleref{ht-rand-tape},  the reference  eventually stores the modular addition of both presampled values at the end, matching the one sampled in $\otpspec$.

It is worth reminding ourselves that the use of presampling tapes in this example is slightly different from that of the batch sampling example in \cref{sec:advanced-rules-foxtrot}, where previously, we used tapes to \emph{combine} $\Rand$ operations, where here, we use them  to \emph{linearize} the presampling operations to simplify the coupling argument.

To the best of our knowledge, the verification of $\otpimpl$ is out of scope of \emph{all} previous techniques due to its combined use of concurrency, probability, and local state; specifically the probabilistic change in the value of the shared reference is extremely difficult, if not impossible, to capture without ghost resources.

\section{Algebraic Theory}
\label{app:algebraic-theory}
We consider the algebraic theory induced by contextual equivalence in \theaplog, summarized in \cref{fig:eq-theory}.

We first define syntactic sugar for binary nondeterministic ($\mathbf{or}$) and probabilistic choice ($\oplus$). To implement nondeterministic choice, we nondeterministically generate a number with $\nondet~\TT$, compare it with $0$, and return $e_1$ or $e_2$ depending on the result of the test. To implement binary probabilistic choice of some rational bias $0\leq p \leq 1$, we sample from $\Rand \tapebound$ and execute $e_1$ or $e_2$ depending on whether the value is smaller than $\tapeboundB$, where $\tapebound$ and $\tapeboundB$ are the smallest integers such that $p=\tapeboundB/(\tapebound+1)$.
  \smalldisplay{  \begin{align*}
      \ndchoice{e_1}{e_2}\eqdef{}&\If \nondet~\TT=0 then e_1 \Else e_2\\
      {e_1}\oplus_p{e_2}\eqdef{}&\If \Rand \tapebound <\tapeboundB then e_1 \Else e_2 \quad \text{where $\tapeboundB/(\tapebound+1)=p$}
    \end{align*}}
    The first three equations in \cref{fig:eq-theory} show that the probabilistic choice operator satisfies the equational theory of a convex algebra. The next four equations show that the nondeterministic choice operator satisfies the equational theory of a join semilattice. 
    \begin{figure}[ht!]
      \small
	\begin{align*}
		e_1 \oplus_{p} e_1 &\ctxeqrel e_1 \\
	e_1 \oplus_{p} e_2 &\ctxeqrel e_2 \oplus_{1-p} e_1 \\
	(e_1 \oplus_{p} e_2) \oplus_{q} e_3 &\ctxeqrel e_1 \oplus_{pq} (e_2 \oplus_{\frac{q-pq}{1-pq}} e_3) \\
	\ndchoice{e_1}{e_1} &\ctxeqrel e_1  \\
	\ndchoice{e_1}{e_2} &\ctxeqrel \ndchoice{e_2}{e_1}  \\
	\ndchoice{e_1}{(\ndchoice{e_2}{e_3})} &\ctxeqrel \ndchoice{(\ndchoice{e_1}{e_2})}{e_3}  \\
		\ndchoice{e_1}{(\diverge~\TT)} &\ctxeqrel e_1  \\
	 \ndchoice{(e_1 \oplus_p e_2)}{(e_1 \oplus_p e_3)}&\ctxrefinessym  e_1 \oplus_p (\ndchoice{e_2}{e_3})
	\end{align*}
	\caption{Equational theory. Here, $\tau\in\Type$ and $\All i\in\{1,2,3\}. \emptyset \vdash e_i:\tau$.}
	\label{fig:eq-theory}
\end{figure}

The last equation is the distributive law of a convex semilattice. Currently,  \theaplog is only expressive enough to prove one direction of the equivalence. We hypothesis that the other direction requires more advanced logical facilities for reasoning about probability and nondeterminism such as prophecy variables~\cite{prophecy1, prophecy2, iris-prophecy} (as it is the case similarly for an equation in ReLoC~\cite{reloc}) and we leave it as future work. 

Almost all of these equations are straightforward to prove; the only exception is that of the third equation, as it involves one to reason about the coupling of a pair of probabilistic choices on both sides, where all four $\Rand$ operations are sampling from \emph{different} ranges. To reason about this equation, we adopt the key ideas captured by \hyperlink{slogan2}{\textbf{Slogan 2}} (see \cref{sec:advanced-rules-foxtrot}), where we introduce several intermediate programs that utilize presampling tapes or parallelize the $\Rand$ operations,  and apply a relatively complex coupling rule that relates all four probabilistic samplings in one go, which we omit for brevity.

\section{Proof Sketch of \cref{thm:adequacy-simpl}}
\label{app:adequacy-proof}
We provide a proof sketch of the intermediate result \cref{thm:adequacy-simpl} from \cref{sec:soundness} that is used to prove the adequacy theorem \cref{thm:adequacy}.

 Firstly, we prove  \cref{lem:arcoupl-exp-r}, which informally states that approximate couplings are composable, enabling us to chain couplings of single steps of executions together. This  result was first presented in Approxis~\cite{approxis}, where the grading $\err'$ can depend on the value of $b$. We can also derive a similar lemma where one varies the error on $A$ instead, but that is not used in the model of \theaplog. 
 \begin{lemma}
   \label[lemma]{lem:arcoupl-exp-r}
   Let \(\Err : B \ra [0,1]\).
   If \(\ARcoupl {\distr_1} {\distr_2} \err R\)
   and \(\forall (a,b) \in R, \ARcoupl {f(a)} {g(b)} {\Err(b)} {R'}\),
   then \(\ARcoupl {(\distr_1 \mbindi f)} {(\distr_2 \mbindi g)} {\err + \err'} {R'}\)
   where \(\err' = \expect[\distr_2]{\Err}\).
 \end{lemma}

 We also define the trivial  FIsch $\initialfisch$ that only returns $\None$:
\begin{lemma}\label{lem:empty-fisch}
  There exists a FIsch $\initialfisch$ such that for all $x$, we have $\limfiexec_{\initialfisch} (x)=\mret x$.
\end{lemma}

We are now ready to provide a proof sketch of \cref{thm:adequacy-simpl}.
\begin{proofsketch}
  The overall proof sketch is similar to a bi-simulation argument.
  We start by taking induction on the number of steps $n$ taken on the implementation side.

  Let us first consider the inductive case.
  The key idea for the inductive case is that for every action that occurs on the implementation side, i.e.~the left-hand side, we extend the  FIsch on the specification side, i.e.~the right-hand side, such that the extended FIsch on the right continues to satisfy the final approximate coupling goal.

  There are four types of actions on the left-hand side to consider during the inductive case:
  \begin{enumerate}
  \item The left scheduler chooses a thread to step and updates its internal scheduler state
  \item The scheduler erasable state update from the first $\specStepp$ is applied to the left configuration
  \item The thread chosen takes a step via the $\progStepp$
  \item The scheduler erasable state update from the second $\specStepp$ is applied to the left configuration 
  \end{enumerate}

  As an example, we explain the bi-simulation argument for the second type of action. (We omit the other cases for brevity.) We start by taking induction on the $\specStepp$ fixpoint, and we  outline the proof sketch for the most challenging case \ruleref{spec-step-exp}. 

  Given the premises of \ruleref{spec-step-exp}, we first rewrite the left-hand side distribution to be $\distr\mbindi(\Lam \state'.\execVal_{\sch,n}(\schstate, (\expr\cons\vec\expr, \state')))$ with the $\scherasable$ condition. It then suffices to find some function $f$ such that $\appfisch(\fisch,f)$ is the oscheduler that satisfies the approximate coupling, where $\fisch$ is the FIsch from the inductive hypothesis. We then apply \cref{lem:app-fisch} and it suffices to prove the approximate coupling for two ``bind'' distributions. We apply \cref{lem:arcoupl-exp-r} and it suffices to show that there is an approximate coupling between $\distr$ and the execution of $\fisch$, which is provided by one of the conditions of \ruleref{spec-step-exp}, and for every intermediate result satisfying the relation $R$, the chosen $\fisch'$ according to the function $f$ satisfies the approximate coupling for the $(\Lam \state'.\execVal_{\sch,n}(\schstate, (\expr\cons\vec\expr, \state')))$ continuation.

    Here instead of providing the explicit function $f$, we apply \cref{thm:choice}, meaning for each intermediate  result that satisfies $R$, we need to find a single FIsch $\fisch'$ that satisfies the approximate coupling. This is done by simply applying the inductive hypothesis of $\specStepp$. Here the $\stateinjective(R)$ condition is essential because the function $f$ can only depend on the FIsch state on the right hand side, not the configuration on the right hand side or resulting state on the left hand side. The $\stateinjective(R)$ condition ensures that the FIsch state $\schstate'$ on the right hand side is unique if there exists states on the left and configurations on the right that satisfies $R$ together with $\schstate'$.

  For the other types of actions, we follow a similar argument as the first $\specStepp$ in extending the FIsch on the right.
  The argument is also the same for the base case of the induction. In particular, after  ``stripping'' away one $\specStepp$, and we ultimately choose $\initialfisch$ (\cref{lem:empty-fisch}) to be the FIsch on the right that approximately couples the distribution on the left. 
  
\end{proofsketch}

\section{Proof of \ruleref{ht-rand}}
\label{app:ht-rand}

As mentioned in \cref{sec:key-ideas}, with the new definition of the weakest precondition and model, we also have to prove all the proof rules of \theaplog from scratch, which is \emph{not} automatic.
As a proof of concept, we provide a proof sketch of the following rule \ruleref{ht-rand} for symbolically executing a $\Rand$ on the left-hand side program without coupling any execution on the right.
\smalldisplay{
\[
  \inferH{ht-rand}
	{  }
	{ \hoare{\TRUE}{\Rand N}{n \ldotp n \in \intrange{0}{N}} } \]}%
To do so, we present another way of constructing a FIsch, which intuitively performs one single step, and depending on the outcome, proceeds according to a function $f$.
\begin{lemma}\label{lem:cons-fisch}
  There exists a function $\consfisch$ such that
  given function  $f:\Cfg'\ra\DDistr(\nat)$ and another function $g:\nat\ra\FIsch$,  $\consfisch(f, g)$ is  a FIsch where for all $\cfg$, we have
  $\limfiexec_{\consfisch(f,g)} (\nil,\cfg)=
  f (\cfg) \mbindi (\Lam \tid. \fitpstepdistr(\cfg)(\tid) \mbindi (h(\tid)))$ 
  where
  \[h(\tid)\eqdef{}
  (\Lam \cfg'.
  \limfiexec_{\liftfisch([(\cfg,\tid)], g(\tid))}([(\cfg,\tid)],\cfg')
  )\]
\end{lemma}

\begin{lemma}
  The rule \ruleref{ht-rand} is sound in the model of \theaplog.
\end{lemma}
\begin{proofsketch}
  After unfolding the definitions of the weakest precondition, it suffices to use the $\progStepp$  \ruleref{prog-step-exp} to prove \ruleref{ht-rand}.

  The tricky part of this proof is that although we are only stepping on the left-hand side program, we would need to provide a FIsch on the right that couples all the possible changes from the $\Rand$ expression without taking \emph{any} steps on the program on the right-hand side. The key idea is to stutter the FIsch on the right by choosing to step threads that are outside the range (recall that $\fitpstepdistr$ returns the same configuration if we choose to step a thread that does not exist). In particular we choose the FIsch $\fisch$ defined as follows for the coupling in $\progStepp$:
  \[\fisch\eqdef{}\consfisch((\Lam \cfg. \unifd \tapebound \mbindi (\Lam \tid. \mret(\tid + \text{length} (\cfg.1)))),(\Lam \_. \initialfisch))\]

  This FIsch $\fisch$ intuitively takes a stutter step by attempting to step the thread $\tid+\text{length}(\cfg.1)$ where $\tid$ is uniformly chosen from the distribution $\unifd \tapebound$. Since the thread chosen to step is added as information to the state of the FIsch,  we can easily construct a coupling between the updated states of $\fisch$ and the result of $\Rand \tapebound$. Moreover, this updated state of $\fisch$ is uniquely defined by the result of $\Rand \tapebound$ so  we can easily define a relation $R$ that is state-injective. 

\end{proofsketch}

}{}

\end{document}